\documentclass[11pt]{article}
\usepackage[dvipsnames,x11names]{xcolor}
\usepackage{epsfig}
\usepackage{lscape}
\usepackage{amssymb} 
\usepackage{graphicx}
\usepackage{footnote}
\usepackage[utf8]{inputenc}	
\usepackage{mathtools}
\usepackage{amsthm}
\usepackage{wasysym}
\usepackage{amsfonts}
\usepackage[english]{babel}
\usepackage{bigints}
\usepackage{mathrsfs}
\usepackage{lineno}
\usepackage{nicefrac}
\usepackage{enumerate}
\usepackage{appendix}
\usepackage{graphicx}
\usepackage{commath}
\usepackage{multirow}
\usepackage{tabu}

\usepackage{times}
\usepackage[T1]{fontenc}

\usepackage{algorithm} 
\usepackage{algpseudocode} 

\usepackage[most]{tcolorbox}
\usepackage{mdframed}

\usepackage{amsmath}
\usepackage{mathtools} 
\usepackage{mathrsfs}

\usepackage[]{natbib}
\usepackage{csquotes}

\usepackage{authblk}

\usepackage[colorlinks=true,citecolor=black,linkcolor=black,urlcolor=blue]{hyperref}
\definecolor{light-gray}{gray}{0.91}

\usepackage[margin=2cm]{geometry}

\allowdisplaybreaks[1]

\makeindex

\usepackage{sectsty}
\sectionfont{\color{BrickRed}}

\sectionfont{\color{DodgerBlue4}}
\subsectionfont{\color{DodgerBlue4}}
\subsubsectionfont{\color{DodgerBlue4}}
\paragraphfont{\color{DodgerBlue4}}

\usepackage{authblk}

\newcommand{\imag}{\textbf{i}}

\def\Lm{\mathsf{L}}

\def\Linop{\mathcal{L}}
\def\adLinop{\mathcal{L}^{\dagger}}

\def\d{\textbf{d}}
\def\u{\textbf{u}}

\def\m{\textbf{m}}
\def\v{\textbf{v}}
\def\x{\textbf{x}}
\def\f{\textbf{f}}

\def\E{\textbf{E}}

\def\WF{\boldsymbol{\psi}}
\def\id{\mathbb{I}}
\def\error{\mathcal{E}}

\usepackage{color}

\usepackage{microtype} 

\begin{document}

\begin{flushleft}
	\flushleft
	\setlength{\baselineskip}{2.3em} 
	\textbf{\huge \color{DodgerBlue4}{On the Fr\'echet interaction density of certain wave equations}}\vspace{1ex}
	
	\vspace{1em}  
	
	\setlength{\baselineskip}{\normalbaselineskip} 
	
	\large{Rafael Abreu}\textsuperscript{1,*} and \large{Chahana Nagesh}\textsuperscript{1} \\[0.5em]  
	
	\scriptsize 
	\textsuperscript{1} \textit{Institut de Physique du Globe de Paris, CNRS, Université de Paris, Paris, France} \\
	\textsuperscript{*} {email: rabreu@ipgp.fr}
	
	\normalsize 
\end{flushleft}

\begin{abstract}
	We extend the adjoint method to complex-valued PDEs and introduce the \emph{Fr\'echet interaction density}, as the most fundamental interaction from which Fr\'echet sensitivity kernels can be derived. We apply this framework to four representative equations: two real-valued PDEs (the second-order wave equation and the Euler--Bernoulli beam equation) and two complex-valued PDEs (the complex transport equation and the Schr\"odinger equation with zero potential). We compute and analyze the Fr\'echet interaction densities for all four PDEs and show that the interaction shows consistent structure, with a waveform that depends on the initial conditions. For the  Schr\"odinger equation, when the adjoint field is chosen as the complex conjugate of the forward wavefunction, the interaction density reduces algebraically to the Born probability density.  Our results establish a unified approach to sensitivity analysis for real- and complex-valued PDEs.
\end{abstract}

\noindent{\footnotesize \textbf{Keywords:} Adjoint method; Fr\'echet kernels; Complex PDEs; Schr\"odinger equation; Born rule}

\newpage



\section{Introduction}

The Adjoint Equation method and the Adjoint State method are closely related mathematical frameworks used to analyze systems governed by partial differential equations (PDEs). The Adjoint Equation refers to the differential equation obtained from the variational, or weak formulation, of a forward problem. It expresses how perturbations in a system's parameters affects the outputs. This is calculated using an adjoint (dual) field that satisfies a specific PDE (adjoint equation) with corresponding boundary/terminal conditions. The adjoint equation is therefore a mathematical construct, derived to quantify sensitivities properties of the forward model \citep[e.g.][]{tonti1973variational}.
 
By contrast, the Adjoint State method is a computational procedure that uses the adjoint equation within an optimization or inverse problem framework. The adjoint field enforces the PDE constraint. Solving both the forward and adjoint problems enables the efficient computation of gradients of the cost functional with respect to model parameters, independently of the number of parameters \citep[e.g.][]{menke2012geophysical,fichtner2010book}. In summary, the Adjoint Equation method emphasizes the derivation of the adjoint equations, while the Adjoint State method highlights the use in optimization/sensitivity problems.

The connection between the adjoint equation and gradient computation naturally leads to the concept of Fr\'echet sensitivity kernels. These Fr\'echet kernels provide a systematic way to quantify how perturbations in model parameters affect observables. In other words, the Fr\'echet sensitivity kernel encapsulates the (linearized) influence of the changes of the PDE parameters on the output. This can be efficiently computed using the solution of the adjoint equation. 

In this work, we generalize the adjoint framework to complex-valued PDEs, extending classical formulations that are typically restricted to real-valued systems. We demonstrate this generalization through four representative cases: (i) two real-valued PDEs—the second-order wave equation and the Euler–Bernoulli beam equation; and (ii) two complex-valued PDEs—the complex transport equation and the Schr\"odinger equation with zero potential.

We make this specific choice because the real-valued PDEs can be viewed as the real components of the complex-valued PDEs. This help to establish a direct correspondence between the two groups. This design allows us to interpret complex-valued PDEs as natural extensions of well-understood real-valued systems.

We introduce the concept of Fr\'echet interaction density, which we define as the most fundamental density from which all Fr\'echet sensitivity kernels can be derived. We compute and analyze the Fr\'echet interaction density and its time integral for each of the four PDEs considered. Interestingly, for the particular case of the Schr\"odinger equation with zero potential and an appropriate choice of the adjoint variable, the Fr\'echet interaction density coincides with the Born rule of quantum mechanics \citep[e.g.][]{griffiths2018introduction,shankar2012principles,zee2010quantum,sakurai2020modern}. This suggests a link between sensitivity analysis and the probabilistic interpretation of wavefunctions.

All PDEs and adjoint problems are solved numerically using the Finite Difference method \citep[][]{moczo2014finite,igel2017computational}. The analysis is carried out in one spatial dimension, which simplifies the problem and captures the essential results while maintaining clarity. We finally discuss the implications of our findings for sensitivity analysis and inverse theory, and outline possible extensions of the complex-valued adjoint framework to multidimensional and nonlinear systems.

\section{Mathematical preliminaries}
\label{sec.notational_agreement}

We briefly review fundamental concepts of functional analysis used in this work. For a more detailed introduction we refer to \cite{zeidler1984nonlinear,zeidler1990nonlinear,griffel2002applied}.

\subsection{Inner products}

We denote by the inner product of two scalar $u(x),v(x)$ and two vector $\u(\x),\v(\x)$ functions, as follows
\begin{align}
	\left\langle u,v \right\rangle_x = \int_x u \, v \dif x, \qquad \left\langle \u,\v \right\rangle_x =  \int_{\Omega} \v^T \u \dif \x =  \int_{\Omega} \u \cdot \v \dif \x = \int_{\Omega} u_i \, v_i \dif \x,
\end{align}
respectively. Analogously, the inner product over time and space of two scalars $u(x,t),v(x,t)$ and two vectors $\u(\x,t),\v(\x,t)$ functions, as follows
\begin{align}
	\left\langle u,v \right\rangle_{x,t} = \int_x \int_T u \, v \dif t \dif x , \qquad \left\langle \u,\v \right\rangle_{x,t} = \int_{\Omega} \int_T \v^T \u \dif t \dif \x =  \int_{\Omega} \int_T \u \cdot \v \dif t \dif \x = \int_{\Omega} \int_T  u_i \, v_i \dif t \dif \x,
\end{align}
respectively. 

A symmetric matrix is defined as a matrix $A$ such that $A^T=A$, where $A^T$ refers to the transposed matrix defined by $A^T_{ij}=A_{ji}$. Then it follows that for any $x,y\in \mathbb{R}^n$,
\begin{align}
	\left\langle x,Ay \right\rangle = \left\langle A^Tx,y \right\rangle .
	\label{eq.Matrix_adjoint}
\end{align} 

\subsection{Bilinear forms}

Let $X$ be a real Banach space (a complete normed vector space). A bilinear form on $X$ is a map $A:X\times X\to \mathbb{R}$ with the following properties \citep{zeidler1990nonlinear} 
\begin{align}
	\begin{aligned}
		\begin{cases}
			A(u,\alpha v + \beta w) & = \alpha A(u,v) + \beta A(u,w) , \\
			A(\alpha v + \beta w,u) & = \alpha A(v,u) + \beta A(w,u) ,
		\end{cases}		
	\end{aligned}
\end{align}
for all $u,v,w\in X$ and all $\alpha,\beta \in \mathbb{R}$.

\subsection{Linear operators and their adjoints}

Let $N,M$ be vector spaces. An operator $\Linop:N\to M$ is linear if 
\begin{align}
	\Linop(ax+by) = a \Linop x + b \Linop y ,
\end{align}
for all scalar $a,b$ and all $x,y\in N$. The adjoint\footnote{the etymology of the word adjoint traces to the Latin \textit{adiungo} which means ad- (``to, towards, at'') + iungo (``join, connect, attach''). } of a linear operator comes from a generalization of the matrix inner product given in eq. \eqref{eq.Matrix_adjoint}. Let $\Linop:\mathcal{H}\to \mathcal{H}$ be a bounded linear operator on a Hilbert space $\mathcal{H}$\footnote{A Hilbert space $X$ is a linear space together with a scalar product with the additional property that each Cauchy sequence is convergent. This means that a Hilbert space defines a distance function (induced by the scalar product) for which the space is a complete metric space (a convergent Cauchy sequence of points in $m$ has a limit that is also in $m$) \citep{zeidler1990nonlinear}. }, then there is a unique operator $\Linop^{\dagger}:\mathcal{H}\to \mathcal{H}$ such that
\begin{align}
	\left\langle x,\Linop^{\dagger}y \right\rangle = \left\langle \Linop x,y \right\rangle \qquad \text{for all} \quad x,y\in \mathcal{H} .
\end{align}
The linear and bounded operator $\adLinop$ is called the adjoint operator of $\Linop$. 

A linear bounded operator $\Linop:N\to M$ is invertible if for each $x\in M$ there is one and only one $y\in N$ such that $\Linop y = x$. The mapping $x \mapsto y$ is called the inverse of $\Linop$ and we denote it by $y=\Linop^{-1}x$. The adjoint operator $\adLinop$ satisfies \citep{griffel2002applied}
\begin{align}
	\left(\Linop^{\dagger} \right)^{\dagger}=\Linop, \qquad  \left(\Linop^{\dagger} \right)^{-1} = \left(\Linop^{-1} \right)^{\dagger} , \qquad \left(\Linop_1 \Linop_2 \right)^{\dagger} = \left(\Linop_2\right)^{\dagger} \left(\Linop_1 \right)^{\dagger} ,
\end{align}
where $\Linop_1,\Linop_2$ refer to two different linear bounded operators. The adjoint operator is called self-adjoint (or Hermitian) if $\adLinop=\Linop$.

\subsection{Adjoint vs convolutional adjoint operators} 

There are two key points to remember about the adjoint operators previously defined:
\begin{enumerate}
	\item The adjoint operator \( A^{\dagger} \) is typically defined with respect to an inner product space.
	\item The adjoint satisfies the condition that the inner products \( \langle A x, y \rangle = \langle x, A^{\dagger} y \rangle, \forall x , y \).
\end{enumerate}

In the context of a Hilbert space \( \mathcal{H} \), the convolution of two functions \( f \) and \( g \) is defined as the inner product between \( f \) and the shifted version of \( g \) as follows:
\begin{align}
	(f \star g)(x) = \int_{-\infty}^{\infty} f(\tau)g(x-\tau) \dif \tau = \langle f(\cdot), g(x - \cdot) \rangle = \langle f , g \rangle^c ,
	\label{eq.convolution}
\end{align}
where \( x \) is the point at which the convolution is evaluated, and \( g(x - \cdot) \) represents the shifted function \( g \), with \( x \) shifting the argument of \( g \). 

The Adjoint operator in terms of the convolution bilinear form is defined as follows: Let \( A \) be a bounded linear operator acting on a Hilbert space \( \mathcal{H} \). The adjoint operator \( A^{\dagger} \) of \( A \) with respect to the convolution bilinear form satisfies the following equation for all \( x, y \in \mathcal{H} \):
\begin{align}
	\langle A x, y \rangle = \langle x, A^{\dagger} y \rangle ^c, \quad \forall x , y \in \mathcal{H},
\end{align}
where \( \langle \cdot, \cdot \rangle^c \) denotes the convolution on the Hilbert space \( \mathcal{H} \). The operator \( A^{\dagger} \) is the unique operator that satisfies this relation, provided that \( A \) is a bounded operator \citep{tonti1973variational}.

\subsection{Differential calculus of operators: Functional/Fr\'echet derivative}

A continuous linear operator $\Linop:N\to M$ is said to be the Fr\'echet derivative of $\boldsymbol{f}:N\to M$ at the point $\textbf{x}\in N$ if \citep{frechet1911notion,frechet1912notion,frechet1925notion,zeidler1984nonlinear,griffel2002applied} 
\begin{align}
	\boldsymbol{f}(\textbf{x}+\delta \boldsymbol{h})=\boldsymbol{f}(\textbf{x}) + \Linop \delta \boldsymbol{h} + o(\delta \boldsymbol{h}) \qquad \text{as} \quad \delta \boldsymbol{h}\to 0,
	\label{eq.General_Frechet_der}
\end{align}
where $o(\delta \boldsymbol{h})$ denotes the set of all functions $\boldsymbol{f}(\delta \boldsymbol{h})=o(\delta \boldsymbol{h})$ that satisfy $\norm{\boldsymbol{f}(\delta \boldsymbol{h})}/\norm{\delta \boldsymbol{h}} \to 0$ as $\delta \boldsymbol{h}\to 0$. 

The expression $\boldsymbol{f}(\delta \boldsymbol{h})=o(\delta \boldsymbol{h})$ means that $\boldsymbol{f}$ is of a smaller order of magnitude than $\delta \boldsymbol{h}$. More precisely, a functional $\boldsymbol{f}$ is called Fr\'echet differentiable if there exists a linear continuous operator $\Linop$ such that 
\begin{align}
	\lim_{\norm{\boldsymbol{h}}\to 0} \frac{\norm{\boldsymbol{f}(\textbf{x}+\delta \boldsymbol{h})-\boldsymbol{f}(\textbf{x})-\Linop \delta \boldsymbol{h}}}{\norm{\boldsymbol{h}}}= 0,
	\label{eq.Frechet}
\end{align}
where $\norm{}$ denotes the operator norm. A variation of any functional $\boldsymbol{h}$ by an infinitesimal but arbitrary amount can be represented in the form
\begin{align}
	\delta \boldsymbol{h} (\textbf{x}) = \epsilon \boldsymbol{\eta}(\textbf{x}),
\end{align}
where $\epsilon$ is an infinitesimal number and $\boldsymbol{\eta}$ is an arbitrary function; we thus can write $\boldsymbol{f}(\textbf{x}+\delta \boldsymbol{h})=\boldsymbol{f}(\textbf{x}+\epsilon\boldsymbol{\eta})$. Considering that $\textbf{x}\in\Omega$, we can write the following
\begin{align}
	\Linop \delta \boldsymbol{h} =  \boldsymbol{f}(\textbf{x}+\delta \boldsymbol{h})-\boldsymbol{f}(\textbf{x}) = \delta \boldsymbol{f} = \int_{\Omega} \left[\frac{\delta \boldsymbol{f}}{\delta \boldsymbol{h}}\right] \delta \boldsymbol{h} \dif \textbf{x} = \epsilon \int_{\Omega} \left[\frac{\delta \boldsymbol{f}}{\delta \boldsymbol{h}}\right] \boldsymbol{\eta} \dif \textbf{x} = \epsilon \int_{\Omega} \Linop \boldsymbol{\eta} \dif \textbf{x}.
	\label{eq.Frechet_int}
\end{align}

This definition of the Fr\'echet derivative implies the form of an integral linear function with kernel $\delta \boldsymbol{f}/\delta \boldsymbol{h}$ acting on a function $\boldsymbol{\eta}$. Note, however, that the definition given in eq. \eqref{eq.Frechet_int} is not guaranteed for arbitrary functionals $\boldsymbol{\eta}$ and $\boldsymbol{f}$. 

The integral inserted in eq. \eqref{eq.Frechet_int} is done by assuming that each small change in the model parameters $\m$ influences the data observed $\d$ at different locations $\x$, and the Fr\'echet derivative kernel $\delta \boldsymbol{f}/ \delta \boldsymbol{h}$ quantifies this. Thus, to find the overall effect of a change $\delta \boldsymbol{f}$ in the model, we sum (integrate) the contributions of this change to the data at all locations.

Consider a (nonlinear) functional $\textbf{G}$ that maps the functional of the model space $\textbf{m}$ to the data space $\textbf{d}$ as follows:
\begin{align}
	\textbf{d} = \textbf{G}(\textbf{m}) .
	\label{eq.Geophysical_data}
\end{align} 
Combining eq. \eqref{eq.Frechet_int} and eq. \eqref{eq.Geophysical_data}, we can write
\begin{align}
	\delta \textbf{d} = \delta \textbf{G} = \int_{\Omega} \left[\frac{\delta \textbf{G}}{\delta \textbf{m}}\right] \delta \textbf{m} \dif \textbf{x} =  \int_{\Omega} \Linop \delta \textbf{m}\dif \textbf{x} ,
	\label{eq.Frechet_geophysical}
\end{align}
where the Fr\'echet derivative kernel in eq. \eqref{eq.Frechet_geophysical} is $\Linop = \partial \textbf{G}/\partial \textbf{m}$. 

\section{The (complex--valued) Adjoint Equation method}
\label{sec.General_Adjoint_method}

We next present a compact complex valued operator formulation of the adjoint equation method. To keep generality in the presentation, we assume that any physical observation that we call data \( \d \in \mathbb{C}^n \) can (accurately enough) be described by a linear differential or integral operator \( \Linop(\m) \), parameterized by certain model parameters \( \m \in \mathcal{M} \subseteq \mathbb{C}^n\), as follows
\begin{align}
	\d = \Linop(\m) \u ,
	\label{eq.linear_operator_data}
\end{align}
where \( \u \in \mathcal{U} \subseteq \mathbb{C}^n\) denotes the dependent variable(s) and
\begin{align}
	\Linop(\m) : \mathcal{U} \rightarrow \mathbb{C}^n
\end{align}
is a linear (differential or integral) operator with complex coefficients, i.e., $\Linop(a f + b g) = a \Linop f + b \Linop g \quad \forall a, b \in \mathbb{C}^n$ acting on functions \( f,g \in\mathcal{U} \subseteq \mathbb{C}^n \).

We aim to minimize the misfit or error function $\E$ defined as any discrepancy between observations $(\d)$ and synthetic data $(\u)$ obtained by modeling the physical event of interest using eq. \eqref{eq.linear_operator_data}. The misfit functional $\E$ can be written as follows 
\begin{align}
		\E & = \error (\d ,\u),
	\label{eq.chi_squared}
\end{align} 
where $\error$ represents an operator that computes the discrepancy between observed $(\d)$ and predicted data $(\u)$. We assume that both observed and synthetic data are functions of space and time, i.e., \( \d = \d(\x, t) \) and \( \u = \u(\x, t) \), with $t\in \mathbb{R}$ and $\x\in \mathbb{R}^3$.

\subsection{Lagrange minimization}

We employ the method of Lagrange multipliers to minimize the error function $\E$ (eq. \eqref{eq.chi_squared}) subject to the constraint that the data can (accurately enough) be described by the linear operator $\Linop(\m) \u$. This constrained optimization problem can be formulated via the augmented functional as follows
\begin{align}
	\chi = \int_{0}^{T} \int_{\Omega}  \left[\error (\d,\u) + \Lm \left(\Linop(\m) \u \right)\right] \dif ^3 \x \dif t ,
	\label{eq.Adjoint_General_Functional}
\end{align}
where \( \Omega \) is the spatial domain, and \( \Lm \) is a vector-valued Lagrange multiplier that enforces the physical constraint. Using the convolution bilinear form (see \citep{tonti1973variational,abreu2024understanding}), we can express eq. \eqref{eq.Adjoint_General_Functional} in a more compact operator form as follows
\begin{align}
\chi = \left\langle \id, \error (\d,\u) \right\rangle_{\x,t}  + \left\langle \Lm ,\Linop(\m) \u \right\rangle^c_{\x,t} ,
\label{eq.operator}
\end{align}
where $\id$ is the identity operator. 

\subsection{Karush–Kuhn–Tucker conditions}

The Karush–Kuhn–Tucker (KKT) conditions are a set of necessary conditions for a solution in a constrained optimization problem to be optimal \citep{hanson1981sufficiency,boyd2004convex,hanson1999invexity}. Thus, to minimize the difference between data subject to the constraint imposed by the forward model $\Linop(\m) \u$, the Karush–Kuhn–Tucker (KKT) conditions must be satisfied \citep{boyd2004convex}. KKT conditions in this case can be written as follows
\begin{align}
\left\{
\begin{aligned}
\partial_\m \chi(\m,\Lm,\u,\d) & = 0 , \\
\partial_{\Lm} \chi(\m,\Lm,\u,\d) & = 0 , \\
\partial_\d \chi(\m,\Lm,\u,\d) & = 0 , \\
\partial_\u \chi(\m,\Lm,\u,\d) & = 0 ,
\end{aligned} \right.
\label{eq.KKT_cond}
\end{align}
which lead to the following explicit expressions \citep{menke2012geophysical}
\begin{align}
	\begin{aligned}
	\partial_\m \chi = \left\langle \Lm,\error_{\m} \Linop \u \right\rangle^c_{\x,t} = 0, \qquad \partial_\Lm \chi = \Linop \u  = 0, \qquad \partial_\u \chi = \error_{\u} + \Linop^{\dagger} \Lm = 0 , \qquad \partial_\d \chi = 0 ,
	\end{aligned}
\end{align}
where the dagger $( \dagger)$ denotes the adjoint operator. Note that imposing $\partial_\Lm \chi = \partial_\u \chi = 0$ implies that $ \left\langle \Lm ,\Linop(\m) \u  \right\rangle^c_{\x,t}=0$, which allow us to write
\begin{align}
	\chi = \E .
\end{align} 
As a consequence, the expression
\begin{align}
	\partial_\m \chi=\partial_\m \E=\left\langle \Lm,\delta_{\m} \Linop \u \right\rangle^c_{\x,t} ,
\end{align}
can be understood as the Fr\'echet derivative of the error or misfit function $\E$ (eq. \eqref{eq.chi_squared}) \citep{menke2012geophysical}. 

\section{Applications to real--valued PDEs}

To illustrate the presented theory, we first apply it to two time-dependent 1D PDEs: (1) the conventional second-order acoustic wave and (2) the Euler-Bernoulli beam equations. Despite the adjoint method has been previously documented in the literature for real-valued PDEs \citep[e.g.][]{fichtner2010book}, the next developments will give us some new fundamentals on a different understanding of Fr\'echet sensitivity kernels.

\subsection{The 1D acoustic wave equation}

We consider the 1D wave equation (for fluids) given by the following expression 
\begin{align}
	\Linop u - f= \rho \partial^2_t u - \lambda \partial_x^2 u - f \delta(x-x^s) =0 , \qquad \text{with} \qquad u(x,0)=\partial_t u(x,0) = 0,
	\label{eq.1D_wave_equation}
\end{align}
where $u=u(x,t)$ is the displacement, $\lambda$ is a Lam\'e parameter, $\rho$ the density, $x_s$ the source location and $f=f(t)$ certain time dependent function.

The physics of eq. \eqref{eq.1D_wave_equation} is well known and documented in the literature \citep[e.g.][]{arfken2011mathematical,farlow1993partial,igel2017computational}. Assuming an initial displacement $u(x,0)=g(x)$ and $f(t)=0$, the analytical solution is given by d'Alembert formula
\begin{align}
	u(x,t) = \frac{1}{2}\left[g(x+ct) + g(x-ct)\right] ,
	\label{eq.analytical_solution1}
\end{align} 
where $c=\sqrt{\lambda/\rho}$. These are two waves traveling in opposite directions with no attenuation nor dispersion. Assuming initial conditions at rest, i.e., $u(x,0)=\partial_t u(x,0) = 0$, and a Dirac source time function $f(t)=\delta(t)$, the analytical solution is given by \citep{igel2017computational}
\begin{align}
	u(x,t) = \frac{1}{2c} H \left(t-\frac{\abs{x}}{c}\right) ,
	\label{eq.second_roder_time_solution}
\end{align}
where $H$ is the Heaviside function. For a general $f(t)\neq \delta(t)\neq0$, the analytical solution can be obtained by convolving eq. \eqref{eq.second_roder_time_solution} with the given $f(t)$ \citep{igel2017computational}.

\subsubsection{The 1D acoustic adjoint equations}
\label{sec.Acoustic_Adjoint_Equations}

We now aim to minimize the difference between some data waveforms $\d$ (obtained from certain experiment) and the synthetic waveforms $\u$ obtained using eq. \eqref{eq.1D_wave_equation}. The main idea is to find a velocity model $\sqrt{\lambda/\rho}$ that minimizes these differences (whatever we choose the word \textit{differences} to be). 

We can express this optimization problem as a constrained Lagrange minimization (eq. \eqref{eq.Adjoint_General_Functional}) given by the following expression
\begin{align}
\chi = \frac{1}{2} \sum_{r} \delta (x-x^r)  \left\langle (d  - u), (d-u) \right \rangle_{x,t} - \left\langle \Lm , \left(\rho \partial^2_t u - \lambda \partial_x^2 u - f  \right) \right \rangle_{x,t} ,
\label{eq.1D_Constrained_Action}
\end{align}
where we aim to minimize the waveforms differences $(d  - u)$ at certain locations $x^r$ and the vector Lagrange multiplier $\Lm =\Lm(x,t)$ (which minimizes the functional $\chi$) remains to be determined. 

Taking the first variation $\delta$ of the $\chi$ functional (eq. \eqref{eq.1D_Constrained_Action}) with respect to model parameters $\m=(\rho,\lambda)$, the force term $f$ and displacement $u$,  and using KKT conditions (eqs. \eqref{eq.KKT_cond}) and assuming that the data are accurately enough described by the wave equation, i.e., $\partial_\Lm \chi = \Linop \u - \f= 0$, we obtain the following expressions
\begin{align}
\begin{aligned}
\delta_\m \chi = &   \left\langle   \Lm , \left[\delta \rho \partial_t ^2 u - \delta \lambda \partial_x^2 u - \delta f \right] \right \rangle_{x,t}  = 0 ,\\
\delta_u \chi = &  \sum_{r} \delta(x-x^r) \left\langle   (d  - u),\delta u  \right \rangle_{x,t}  - \left\langle  \Lm, \left[\rho \partial_t ^2 \delta u - \lambda \partial_x^2 u\right] \right \rangle_{x,t} = 0 .
\end{aligned}
\label{eq.1D_General_Constrained_Action}
\end{align}
Integrating by parts (twice) the terms involving spatial and temporal derivatives of the displacement $u$ and the variation $\delta u$, we obtain for $\delta_u \chi$ the following expression
\begin{align}
\begin{aligned}
\delta_u \chi = & \sum_{r} \delta(x-x^r) \left\langle   (d  - u),\delta u  \right \rangle_{x,t} - \left\langle \left[\rho \partial^2_t \Lm - \lambda \partial_x^2 \Lm \right], \delta u\right \rangle_{x,t} = 0 , \\& \text{with} \qquad \Lm(x,0)=\partial_t \Lm(x,0) = 0 .
\end{aligned}
\label{eq.1D_variation_action_integration_parts} 
\end{align}
Here, we assume that $\Lm=\Lm(x,T-t)$ which allows to properly satisfy boundary conditions (see \cite{abreu2024understanding}). In the absence of perturbations in the model parameters $(\delta \rho = \delta \lambda = \delta f = 0)$, i.e., $\delta_\m \chi =0$, the variation in the action $\delta \chi$, reduces to
\begin{align}
\begin{aligned}
\delta \chi = \delta_u \chi = &  \left\langle   \left\{\sum_{r} \delta(x-x^r) (d  - u) - \left[\rho \partial^2_t \Lm -  \lambda \partial_x^2 \Lm \right]\right\}, \delta u\right \rangle_{x,t}  =  0, \\
& \text{with} \quad \Lm(x,0)=\partial_t \Lm(x,0) = 0.
\end{aligned}
\label{eq.1D_reduced_variation_action_integration_parts} 
\end{align}
Next, it follows that the variation in the action $\chi$ is stationary $(\delta_\m \chi + \delta_{\Lm} \chi + \delta_u \chi =0)$ if the following condition holds
\begin{flalign}
\rho \partial^2_t \Lm -  \lambda \partial_x^2 \Lm  = \sum_{r}  \delta (x^r - x)(d  - u) , \qquad \text{with} \qquad \Lm(x,0)=\partial_t \Lm(x,0) = 0 .
\label{eq.1D_adjoint_equation_with_lambda}
\end{flalign}
Note that the expression involving the Lagrange multiplier $(\Lm)$ is equal to the initial wave equation (eq. \eqref{eq.1D_wave_equation}) with a different source term, and that without having previous information on the adjoint wavefield, it seems reasonable to assume
\begin{flalign}
\Lm (x,T-t) \stackrel{\text{def}}{=} u (x,T-t) = u^{\dagger} (x,t),
\label{eq.Reversed_displacement_adjoint}
\end{flalign}
thus, the adjoint wavefield $u^{\dagger}$ is equal to the time-reversed wavefield $u (x,T-t)$ \citep[e.g.][]{Morse1953}. Note also that we could have defined the adjoint wavefield in any other way
\begin{align}
\Lm (x,T-t)\stackrel{\text{def}}{=} \mathcal{F} u(x,T-t) = u^{\dagger} (x,t)
\label{eq.Adjoint_displacement_adjoint}
\end{align}
where $\mathcal{F}$ is any certain linear functional, and the properties previously explained will still hold. For example, we could have rightfully chosen $\Lm(x,T-t)\stackrel{\text{def}}{=} \partial_t u(x,T-t)$. 

The new defined adjoint wavefield must satisfy the equations of motion given for $\Lm$ (eq. \eqref{eq.1D_adjoint_equation_with_lambda}) as follows
\begin{align}
\rho \partial^2_t u^{\dagger} - \lambda \partial_x^2 u^{\dagger}    = \sum_{r}  \delta (x^r - x) (d  - u) . 
\label{eq.1D_elastic_adjoint_wave_equation}
\end{align}
where the adjoint source term is determined by the difference between the data and the synthetic wavefield, \( (d - u) \), at the measurement locations \( \sum_{r} \delta(x^r - x) \).

\subsubsection{1D Fr\'echet derivatives}

In the next, the dependence of the adjoint variable $u^{\dagger}$ on the model is ignored, i.e., we assume that the adjoint wavefield $u^{\dagger}$ perfectly satisfies the adjoint equation of motion (eq. \eqref{eq.1D_adjoint_equation_with_lambda}) and perturbations in the model parameters exist $(\delta \rho \neq \delta \lambda \neq \delta f \neq 0)$. 

\paragraph{The material parameter and source Fr\'echet kernels}

It follows that we can write the variation in the action $\chi$ (eq. \eqref{eq.1D_General_Constrained_Action}) as 
\begin{align}
\begin{aligned}
\delta_\m \chi &= \left\langle u^{\dagger},  \left[\delta \rho \partial_t ^2 u - \delta \lambda \partial_x^2 u - \delta f\right]\right \rangle^c_{x,t} , \\
& = \left\langle  \delta \rho  , u^{\dagger}  \partial_t ^2 u \right \rangle^c_{x,t} - \left\langle \delta \lambda , \partial_x u^{\dagger} \partial_x u \right \rangle^c_{x,t} - \left\langle u^{\dagger} ,\delta f \right \rangle^c_{x,t}, \\
& = \left\langle \delta \ln \rho , K_{\rho} \right \rangle_{x} + \left\langle \delta \ln \lambda , \text{K}_{\lambda} \right \rangle_{x} + \left\langle \delta \ln f , K_{f} \right \rangle_{x} = 0, 
\end{aligned}
\label{eq.1D_elastic_variation_action_kernels}
\end{align}
where we have defined the following sensitivity kernels
\begin{align}
K_{\rho}  = \rho \left\langle  u^{\dagger} , \partial^2_t u  \right \rangle^c_{t} =  - \rho \left\langle   \partial_t u^{\dagger} , \partial_t u \right \rangle_{t},\qquad  \text{K}_{\lambda} =  -\lambda  \left\langle \partial_x  u^{\dagger} , \partial_x u \right \rangle^c_{t}, \qquad K_{f}  = \left\langle  u^{\dagger} ,  f \right \rangle^c_{t} .
\label{eq.1D_Adjoint_Kernels}
\end{align}

Equation \eqref{eq.1D_elastic_variation_action_kernels} shows that the variation of the misfit is a linear functional of the model perturbations. A useful interpretation follows by considering a localized perturbation of the model parameters. 
Choosing, for instance, a point perturbation in density
\begin{align}
	\delta \ln \rho(x) = \delta(x - x_0), \quad \delta \ln \lambda(x)=0, \quad \delta \ln f(x)=0,
\end{align}
we obtain
\begin{align}
	\delta \chi = K_\rho(x_0).
\end{align}
The kernel $K_\rho(x_0)$ represents the change in the misfit caused by a unit perturbation of the model at location $x=x_0$. In this sense, each point $x$ corresponds to a basis direction in model space, and the adjoint method evaluates the misfit sensitivity with respect to all such directions simultaneously. We can build similar interpretations for $K_\lambda$ and $K_f$.

The kernels given in eqs. \eqref{eq.1D_Adjoint_Kernels} describe the change in the misfit function due to changes in the model parameters $(\rho,\lambda)$ and source $(f)$, in terms of the original $(u)$ and adjoint $(u^{\dagger})$ wavefields. These are well known expressions used within the context of geophysical full-waveform inversion  \citep[e.g.][]{fichtner2010book}.

\paragraph{The Fr\'echet interaction density} We define the Fr\'echet interaction density as follows
\begin{align}
	K_R =  u \, u^\dagger  ,
	\label{eq.Frechet_interaction}
\end{align}
which represents the interaction between the forward and adjoint wavefields, and its corresponding time integrated kernel
\begin{align}
\mathsf{K}_R	=\int_T K_R \, \dif t
\end{align} 

We denote by $\Linop_\rho$, $\Linop_\lambda$, and $\Linop_f$ the differential operators related to the model parameters $\rho$, $\lambda$, and $f$, respectively (see eq. \eqref{eq.1D_Adjoint_Kernels}). The Fr\'echet kernels in eq. \eqref{eq.1D_Adjoint_Kernels} can then be written in the general form as follows
\begin{align}
	K_\m  =  \left\langle u^\dagger ,\Linop_\m[u]  \right \rangle^c_{t}, \qquad \text{with} \quad \m \in \{\rho,\lambda,f\} .
\end{align}

The interaction density $K_R(x,t)$ (eq. \eqref{eq.Frechet_interaction}) works as a fundamental bilinear quantity, the instantaneous overlap between forward and adjoint wavefields, from which the Fr\'echet kernels are constructed, and the operators $\mathcal{L}_m$ encode how the governing physics weights this interaction before to the time integration. The physically relevant components of each kernel simply correspond to the derivatives of the wavefield that enter the governing equation, ensuring that the sensitivity reflects how model parameters influence the physics of the PDE in study.

\subsection{The Euler-Bernoulli beam equation}

The Euler-Bernoulli Beam equation given by the following expression
\begin{align}
	\Linop u + V=  \rho A \partial_t^2 u + \partial_x^2 \left(EI \partial_x^2 u\right) + V  \delta(x-x^s) = 0 ,
	\label{eq.Euler_Bernoulli_Beam}
\end{align}
where $u=u(x,t)$, is the transverse displacement, $\rho$ is the mass density, $A$ is the cross-sectional area of the beam, $E$ is the Young's modulus, $I$ is the area moment of inertia,  $x^s$ the source location and $V=V(t)$ is an external time-dependent force \citep{Timoshenko_1951aa,karnopp2012system}. 

\subsubsection{The 1D adjoint Euler-Bernoulli beam equation}

Once again, we assume that we would like to minimize the difference between some data $\d$ and the synthetic waveforms $\u$ (obtained using eq. \eqref{eq.Euler_Bernoulli_Beam}). We express this optimization problem as a constrained Lagrange minimization problem (eq. \eqref{eq.Adjoint_General_Functional}), given by the following expression:
\begin{align}
	\chi = \sum_{r} \delta (x-x^r)  \frac{1}{2}  \left\langle(d  - u),(d  - u) \right \rangle_{x,t} + \left\langle \Lm ,\left[\rho A \partial_t^2 u + \partial_x^2 \left(EI \partial_x^2 u\right) + V \delta(x-x^s) \right]\right \rangle_{x,t} ,
	\label{eq.1D_Euler_Beam_Constrained_Action}
\end{align}
where we aim to minimize the waveforms differences $(d  - u)$ at certain locations $x^r$ and the vector Lagrange multiplier $\Lm(x,t)$ (which minimizes the functional $\chi$) remains to be determined. 

Taking the first variation $\delta$ of the $\chi$ functional (eq. \eqref{eq.1D_Euler_Beam_Constrained_Action}) with respect to model parameters $\m=(\rho A,EI)$, the force term $V$ and displacement $u$,  and using KKT conditions (eqs. \eqref{eq.KKT_cond}) and assuming that the data are accurately enough described by the wave equation and following the detailed procedure described in Sec. \ref{sec.Acoustic_Adjoint_Equations}, leads to the following adjoint equations
\begin{align}
	\rho A \partial^2_t u^{\dagger} + \partial_x^2 \left(EI \partial_x^2 u^{\dagger}\right)    = \sum_{r}  \delta (x^r - x) (d  - u) , 
	\label{eq.1D_Euler_Beam__adjoint_wave_equation}
\end{align}
where the adjoint source term is determined by the difference between the data and the synthetic wavefield, \( (d - u) \), at the measurement locations \( \sum_{r} \delta(x^r - x) \). Note that, for simplification purposes, we have considered the material parameters $(\rho A,EI)$ as single parameters.

\subsubsection{1D Fr\'echet derivatives}

If we assume that the adjoint wavefield $u^{\dagger}$ perfectly satisfies the adjoint equation of motion (eq. \eqref{eq.1D_Euler_Beam__adjoint_wave_equation}) and perturbations in the model parameters exist, we can find expression for the different Fr\'echet kernels.

\paragraph{The material parameter and source Fr\'echet kernels}

It follows that we can write the variation in the action $\chi$ (eq. \eqref{eq.1D_Euler_Beam_Constrained_Action}) as follows 
\begin{align}
	\begin{aligned}
		\delta_\m \chi = \left\langle \delta \ln \rho A , K_{\rho A} \right\rangle_{x} + \left\langle \delta \ln EI , \text{K}_{EI} \right\rangle_{x}+ \left\langle \delta \ln V , K_{V}\right\rangle_{x}  = 0, 
	\end{aligned}
	\label{eq.1D_Euler_Bernoulli_variation_action_kernels}
\end{align}
where we have defined the following sensitivity kernels
\begin{align}
	K_{\rho A}  =\rho A \left\langle \partial_t u^{\dagger} , \partial_t u \right\rangle^c_t,\qquad  \text{K}_{EI} = EI \left\langle \partial_x^2  u^{\dagger} , \partial_x^2 u \right\rangle^c_t, \qquad K_{V}  = \left\langle u^{\dagger} , V \right\rangle^c_t .
	\label{eq.1D_Euler_Bernoulli_Adjoint_Kernels}
\end{align}

The kernels given in eqs. \eqref{eq.1D_Adjoint_Kernels} describe the change in the misfit function due to changes in the model parameters $(\rho A,EI)$ and source $(V)$, in terms of the original $(u)$ and adjoint $(u^{\dagger})$ wavefields. 

\paragraph{The Fr\'echet interaction density} If we denote by $\Linop_{\rho  A}$, $\Linop_{EI}$, and $\Linop_V$ the differential operators related to the model parameters $\rho A$, $EI$, and $V$, respectively (see eq. \eqref{eq.1D_Euler_Bernoulli_Adjoint_Kernels}). The Fr\'echet kernels in eq. \eqref{eq.1D_Euler_Bernoulli_Adjoint_Kernels} can then be written in the general form as follows
\begin{align}
		K_\m  =  \left\langle u^\dagger ,\Linop_\m[u]  \right \rangle^c_{t}, \qquad \text{with} \quad \m \in \{\rho A,E I ,V\} .
	\label{eq.Euler_Bernoulli_General_Frechet}
\end{align}

We can, once again, observe that the interaction density $K_R(x,t)$ (eq. \eqref{eq.Frechet_interaction}) works as a fundamental bilinear quantity, the instantaneous overlap between forward and adjoint wavefields, from which the Fr\'echet kernels defined in eq. \eqref{eq.Euler_Bernoulli_General_Frechet} are constructed. 

\section{Applications to complex--valued PDEs}

We next extend the presented theory to two time-dependent 1D complex PDEs: (1) the complex transport equation and (2) the zero potential Schr\"odinger equation. 

\subsection{The 1D complex transport equation}

We now consider that the displacement $\u$ is a complex valued vector, $(\u\in\mathbb{C})$ and it is governed by the complex 1D transport equation given by the following expression
\begin{align}
	\Linop \, \u - \f =  \imag \sqrt{\rho} \, \partial_t \u - \sqrt{\lambda} \,  \partial_x \bar{\u} - \f \, \delta(x-x^s) =0 , \qquad \text{with} \qquad \u(x,0)=\partial_t \u(x,0) = 0,
	\label{eq.1D_imaginary_wave_equation}
\end{align}
where $\bar{\u}$ is the complex conjugate of $\u$, $\lambda$ is a Lam\'e parameter, $\rho$ the density, $x^s$ the source location and $f=f(t)$ certain time dependent function. The spatial $x$ and temporal $t$ coordinates are considered to be real-valued variables as well as the propagation velocity $c=\sqrt{\lambda/\rho}$, i.e., $\left[x,t,\lambda,\rho\right]\in \mathbb{R}$.

As we will next see, the physics of eq. \eqref{eq.1D_imaginary_wave_equation} is similar to the second-order wave equation (eq. \eqref{eq.1D_wave_equation}) with a slight difference. To illustrate this, we write the complex displacement vector $\u$ as 
\begin{align}
	\u = u_R + \imag \, u_I,
\end{align}
where $u_R,u_I$ are the real and imaginary parts respectively. Assuming constant density $\rho$, we can write eq. \eqref{eq.1D_imaginary_wave_equation}, after a little algebra, as follows
\begin{align}
 \rho \, \partial_t^2 u_R - \lambda \, \partial_x^2 u_R & = f_R \, \delta(x - x^s), \\
 \rho \, \partial_t^2 u_I - \lambda \, \partial_x^2 u_I & = f_I \, \delta(x - x^s),
	\label{eq.1D_complex_wave_equation}
\end{align}
where \( f_R \) and \( f_I \) are the real and imaginary parts of the time dependent source term \( f \), and we have assumed initial conditions at rest, ie.e, $\u(x,0)=\partial_t \u(x,0) = 0$. Note that the real and imaginary parts of the displacement $(u_R,u_I)$ are completely decoupled, and the physics governing each of them is simply the same as the second-order wave equation described before.

\subsubsection{The 1D adjoint complex transport equations}

We can write the constrained Lagrange minimization problem eq. \eqref{eq.Adjoint_General_Functional} as follows
\begin{align}
	\chi = \sum_{r} \delta (x-x^r)  \frac{1}{2}  \left \langle  (\d  - \u), (\d  - \u) \right \rangle_{x,t}  + \left \langle \Lm , \left(\sqrt{\rho} \, \imag \, \partial_t \u - \sqrt{\lambda} \, \partial_x \bar{\u} - \f  \right) \right \rangle_{x,t}   ,
	\label{eq.1D_complex_transport_Constrained_Action}
\end{align}
where the complex vector Lagrange multiplier $\Lm(x,t)$ (which minimizes the functional $\chi$) remains to be determined. 

Taking the first variation of the $\chi$ functional given in eq. \eqref{eq.1D_complex_transport_Constrained_Action} with respect to model parameters $\m=(\rho,\lambda)$, the force term $f$ and displacement $u$,  and using KKT conditions eq. \eqref{eq.KKT_cond} and assuming that the data are accurately enough described by the complex transport equation (eq. \eqref{eq.1D_imaginary_wave_equation}), i.e., $\partial_\Lm \chi = \Linop \u - \f= 0$, we can write the following
\begin{align}
	\begin{aligned}
		\delta_\m \chi = &  \left \langle   \Lm , \left[\frac{1}{2\sqrt{\rho}}\delta \rho \, \imag \,  \partial_t  \u - \frac{1}{2\sqrt{\lambda}} \, \delta \lambda \, \partial_x \bar{\u} - \delta \f \right] \right \rangle_{x,t} = 0 ,\\
		\delta_u \chi = &  \sum_{r} \delta(x-x^r)  \left \langle   (\d  - \u) , \delta \u \right \rangle_{x,t}   -\left \langle  \Lm , \left[\sqrt{\rho} \, \imag \, \partial_t \delta \u - \sqrt{\lambda} \,  \partial_x \delta \bar{\u}\right] \right \rangle_{x,t} = 0 .
	\end{aligned}
	\label{eq.1D_General_Complex_Transport_Constrained_Action}
\end{align}
In the absence of perturbations in the model parameters $(\delta \rho = \delta \lambda = \delta \f = 0)$, i.e., $\delta_\m \chi =0$, and following the same procedure as before, the variation in the action $\chi$ is stationary $(\delta_\m \chi + \delta_{\Lm} \chi + \delta_{\u} \chi =0)$ if
\begin{flalign}
	\sqrt{\rho} \, \imag \, \partial_t \Lm -  \sqrt{\lambda} \, \partial_x \Lm  = \sum_{r}  \delta (x^r - x)(\d  - \u) , \qquad \text{with} \qquad \Lm(x,0) =\partial_t \Lm(x,0) = 0 ,
	\label{eq.1D_complex_transport_adjoint_equation_with_lambda}
\end{flalign}
where again $\Lm=\Lm(x,T-t)$. Without having previous information of the Lagrange multiplier $\Lm$, it seems reasonable to assume $\Lm (x,T-t) \stackrel{\text{def}}{=} \u (x,T-t) = \u^{\dagger} (x,t)$ (eq. \eqref{eq.Reversed_displacement_adjoint}), thus, the adjoint wavefield $\u^{\dagger}$ is equal to the time-reversed complex wavefield $\u (x,T-t)$. The new defined adjoint wavefield must satisfy the equations of motion given for $\Lm$ (eq. \eqref{eq.1D_complex_transport_adjoint_equation_with_lambda}), that is,
\begin{align}
	\sqrt{\rho} \, \imag \, \partial_t  \u^{\dagger} - \sqrt{\lambda} \, \partial_x \bar{\u}^{\dagger} = \sum_{r}  \delta (x^r - x) (\d  - \u) . 
	\label{eq.1D_complex_transport_adjoint_wave_equation}
\end{align}

\subsubsection{1D Fr\'echet derivatives}

In the following, the dependence of $u^{\dagger}$ on the model is ignored, i.e., we assume that the adjoint wavefield $u^{\dagger}$ perfectly satisfies the adjoint equation of motion eq. \eqref{eq.1D_complex_transport_adjoint_equation_with_lambda} and there are perturbations in the model parameters $(\delta \rho \neq \delta \lambda \neq \delta f \neq 0)$. 

\paragraph{The material parameter and source Fr\'echet kernels}

We can write the variation in the action eq. \eqref{eq.1D_General_Complex_Transport_Constrained_Action} as follows 
\begin{align}
	\begin{aligned}
		\delta_\m \chi &= \left \langle \u^{\dagger} ,  \left[\frac{1}{2\sqrt{\rho}}\delta \rho \, \imag \,  \partial_t  \u - \frac{1}{2\sqrt{\lambda}} \delta \lambda \, \partial_x \bar{\u} - \delta \f \right] \right \rangle_{x,t}, \\
		& =  \left \langle  \delta \ln \rho , K_{\rho} \right \rangle_{x} +  \left \langle \delta \ln \lambda , \text{K}_{\lambda} \right \rangle_{x} +  \left \langle \delta \ln \f , K_{\f} \right \rangle_{x} = 0, 
	\end{aligned}
	\label{eq.1D_complex_transport_variation_action_kernels}
\end{align}
where we have defined the following sensitivity kernels
\begin{align}
	K_{\rho}  = \frac{1}{2} \left \langle \sqrt{\rho}\,  \u^{\dagger} ,  \imag \, \partial_t \u \right \rangle_t , \qquad  \text{K}_{\lambda} = - \frac{1}{2} \left \langle \sqrt{\lambda}   \u^{\dagger} , \partial_x \bar{\u} \right \rangle_t , \qquad K_{\f}  = - \left \langle \u^{\dagger} , \f \right \rangle_t .
	\label{eq.1D_complex_transport_Adjoint_Kernels}
\end{align}
Once again, the kernels given in eqs. \eqref{eq.1D_complex_transport_Adjoint_Kernels} give us the change in the misfit function due to changes in the model parameters $(\rho,\lambda)$ and source $(\f)$, in terms of the original $(\u)$ and adjoint $(\u^{\dagger})$ wavefields. Note that we can change the position of the spatial derivative in the expression for $\text{K}_{\lambda}$. By placing the spatial derivative on the Fr\'echet kernel $\text{K}_{\lambda}$, we change the nature of the contribution to the kernel: instead of directly multiplying the wavefield $u^{\dagger}$ with the spatial derivative of $\bar{u}$, we are considering how $\rho$ itself changes in space and how those changes affect the sensitivity $\text{K}_{\lambda}$.

\paragraph{The Fr\'echet interaction density} We define the Fr\'echet interaction density as follows
\begin{align}
	K_R  =   \u \,\u^\dagger  ,
	\label{eq.Complex_Frechet_interaction}
\end{align}
which represents the interaction between the complex forward and adjoint wavefields. 

We denote by $\Linop_\rho$, $\Linop_\lambda$, and $\Linop_\f$ the differential operators related to the model parameters $\rho$, $\lambda$, and $f$, respectively (see eq. \eqref{eq.1D_imaginary_wave_equation}). The Fr\'echet kernels in eq. \eqref{eq.1D_imaginary_wave_equation} can then be written in the general form as follows
\begin{align}
	K_\m =  \left\langle u^\dagger ,\Linop_\m[u]  \right \rangle^c_{t}, \qquad \text{with} \quad \m \in \{\rho,\lambda,\f\} .
\end{align}

We can, once again, observe that the interaction density $K_R(x,t)$ (eq. \eqref{eq.Complex_Frechet_interaction}) works as a fundamental bilinear quantity, the instantaneous overlap between forward and adjoint wavefields, from which the Fr\'echet kernels defined in eq. \eqref{eq.1D_complex_transport_Adjoint_Kernels} are constructed. 

\subsection{The (zero potential) Schr\"odinger equation}

The time-dependent Schr\"odinger equation (TDSE) for a free particle in one dimension (with no potential) reads as follows
\begin{align}
	\Linop \WF + J = \imag \, \hbar \,  \partial_t \WF +\frac{\hbar^2}{2m} \partial^2_x \WF + J = 0 , 
	\label{eq.1D_Schroedinger_equation}
\end{align}
where $\WF(x,t)$ is the complex-valued wavefunction, $\hbar$ is the reduced Planck constant, $m=m(x)$ is the particle effective mass and $J(x,t)$ is a source term that may represent an external force or interaction acting on the particle \citep{zee2010quantum,shankar2012principles,griffiths2018introduction,sakurai2020modern}. 

Note that we have considered a variable mass $(m=m(x))$, which accounts for how the kinetic energy operator changes from one material to another.
If we split the wavefunction $\WF$ into its real ($\psi_R$) and imaginary ($\psi_I$) parts we can write
\begin{align}
	\WF = \psi_R + \imag \psi _I,
\end{align}
and assume $J=0$, after some algebra, one can show that if we decouple real and imaginary parts, the Schr\"odinger equation (eq. \eqref{eq.1D_Schroedinger_equation}) can be written as follows 
\begin{align}
	\left(\partial_t^2 + \frac{\hbar^2}{4m^2} \partial_x ^4 \right) (\psi_R,\psi_I) = 0 .
\end{align}

One can clearly see that the decoupled Schr\"odinger equation has the same mathematical structure as the Euler-Bernoulli Beam equation (eq. \eqref{eq.Euler_Bernoulli_Beam}). 

\subsubsection{The 1D adjoint (zero potential) Schr\"odinger equation}

Using the zero potential adjoint Schr\"odinger equation (eq. \eqref{eq.1D_Schroedinger_equation}), we can write the constrained Lagrange minimization problem eq. \eqref{eq.Adjoint_General_Functional} as follows
\begin{align}
	\chi = \sum_{r} \delta (x-x^r)  \frac{1}{2} \left \langle  (\d  - \WF), (\d  - \WF) \right \rangle_{x,t} -  \left \langle \Lm , \left[ \imag \hbar \partial_t \WF +\frac{\hbar^2}{2m} \partial^2_x \WF + J  \right] \right \rangle_{x,t} ,
	\label{eq.1D_Schroedinger_Constrained_Action}
\end{align}
where the vector Lagrange multiplier $\Lm(x,t)$ (which minimizes the functional $\chi$) remains to be determined. 

Taking the first variation of the $\chi$ functional given in eq. \eqref{eq.1D_Schroedinger_Constrained_Action} with respect to model parameters $\m=(m)$, the force term $J$ and the wavefunction $\WF$, and using KKT conditions eq. \eqref{eq.KKT_cond} and assuming that the data are accurately enough described by the zero potential Schr\"odinger equation (eq. \eqref{eq.1D_Schroedinger_equation}), we can write the following
\begin{align}
	\begin{aligned}
		\delta_\m \chi = &  \left \langle  \Lm , \left[\imag \, \delta \hbar \, \partial_t \WF + \frac{\hbar^2}{2m^2} \delta m \, \partial^2_x \WF - \delta J \right] \right \rangle_{x,t} = 0 ,\\
		\delta_{\psi} \chi = & \left \langle   \sum_{r} \delta(x-x^r) (\d  - \WF) , \delta \WF  \right \rangle_{x,t}  - \left \langle  \Lm , \left[ \imag \, \hbar \, \partial_t \delta \WF +\frac{\hbar^2}{2m} \partial^2_x \delta \WF\right] \right \rangle_{x,t}= 0 .
	\end{aligned}
	\label{eq.1D_General_Complex_Schroedinger_Constrained_Action}
\end{align}

In the absence of perturbations in the model parameters $(\delta m = \delta J = 0)$, i.e., $\delta_\m \chi =0$, and following the same procedure as before, the variation in the action $\chi$ is stationary $(\delta_\m \chi + \delta_{\Lm} \chi + \delta_{\psi} \chi =0)$ if
\begin{flalign}
	\imag \, \hbar \,  \partial_t \WF^{\dagger} +\frac{\hbar^2}{2m} \partial^2_x \WF^{\dagger}  = \sum_{r}  \delta (x^r - x)(\d  - \WF) .
	\label{eq.1D_Schroedinger_adjoint_equation}
\end{flalign}

\subsubsection{1D Fr\'echet derivatives}

In the following, the dependence of $\WF^{\dagger}$ on the model is ignored, i.e., we assume that the wavefunction $\WF^{\dagger}$ perfectly satisfies the adjoint equation of motion eq. \eqref{eq.1D_Schroedinger_adjoint_equation} and there are perturbations in the model parameters $(\delta \hbar \neq \delta m \neq \delta J \neq 0)$. Note that a change in $\hbar$ affects the spread of the wavefunction and the energy levels and the spatial distribution of the wavefunction are directly affected by $\hbar$.

\paragraph{The material parameter and source Fr\'echet kernels}

We can write the variation in the action eq. \eqref{eq.1D_General_Complex_Schroedinger_Constrained_Action} as follows 
\begin{align}
	\begin{aligned}
		\delta_\m \chi &= \left \langle  \WF^{\dagger} , \left[\imag \, \delta \hbar \, \partial_t \WF + \frac{\hbar^2}{2m^2} \delta m \, \partial^2_x \WF + \delta J \right] \right \rangle^c_{x,t} , \\
		& = \left \langle \delta \ln \hbar , K_{\hbar} \right \rangle_{x} + \left \langle \delta \ln m , K_{m} \right \rangle_{x} + \left \langle \delta \ln J , K_{J} \right \rangle_{x} = 0, 
	\end{aligned}
	\label{eq.1D_Schroedinger_variation_action_kernels}
\end{align}
where we have defined the following sensitivity kernels
\begin{align}
K_{\hbar}  = \imag \, \hbar \left \langle  \WF^{\dagger} , \partial_t\WF \right \rangle^c_{t} , \qquad 	K_{m}  = \frac{\hbar^2}{2m}\left \langle  \WF^{\dagger} , \partial_x^2\WF \right \rangle^c_{t} , \qquad K_{J}  = \left \langle \WF^{\dagger} , J \right \rangle^c_{t}.
	\label{eq.1D_Schroedinger_Adjoint_Kernels}
\end{align}

\paragraph{The Fr\'echet interaction density} We define the Fr\'echet interaction density as follows
\begin{align}
	K_R  = \WF \,\WF^\dagger ,
	\label{eq.Schroedinger_Frechet_interaction}
\end{align}
which represents the interaction between the complex forward and adjoint wavefields. 

We denote by $\Linop_{\hbar}$, $\Linop_m$, and $\Linop_J$ the differential operators related to the model parameters $\hbar$, $m$, and $J$, respectively (see eq. \eqref{eq.1D_Schroedinger_equation}). The Fr\'echet kernels in eq. \eqref{eq.1D_Schroedinger_Adjoint_Kernels} can then be written in the general form as follows
\begin{align}
	K_\m = \left \langle \WF^\dagger ,\mathcal{L}_m[\WF] \right \rangle_{t}^c, \qquad \text{with} \quad \m \in \{\hbar,m,J\} .
\end{align}

We can, once again, observe that the interaction density $K_R(x,t)$ (eq. \eqref{eq.Schroedinger_Frechet_interaction}) works as a fundamental bilinear quantity, the instantaneous overlap between forward and adjoint wavefields, from which the Fr\'echet kernels defined in eq. \eqref{eq.1D_Schroedinger_Adjoint_Kernels} are constructed.

\section{Fr\'echet interaction density gallery}

We next analyze the Fr\'echet interaction densities previously obtained for each PDE. For the numerical solution of all  wave equations, we apply the finite-difference method on uniform grids (see \cite{moczo2014finite,igel2017computational} for further details). The main objective of the numerical experiments is to compute and compare the resulting forward–adjoint interaction densities in homogeneous media.

\subsection{Selection of the source time functions} 

We assume initial conditions at rest and in order to generate motion we consider a time dependent source time function $f(t)$ at certain locations $x=x_0$. Two cases for $f(t)$ are assumed:
\begin{enumerate}
	\item We assume that $f(t)$ is a real valued function, i.e., $f\in \mathbb{R}$, and given by a Gaussian pulse centered at $t=t_0$ with certain dominant frequency $f_0$ and amplitude $A$, given by the following expression
	\begin{align}
		f(t) = A \exp \left(-\omega_0^2\frac{(t-t_0)^2}{2}\right) , \quad \text{with} \quad \omega_0 = 2 \pi f_0 .
		\label{eq.source_time_function_Gaussian}
	\end{align}
	
	\item We assume that $f(t)$ is a complex valued function, i.e., $f\in \mathbb{C}$, and given by a modulated Gaussian pulse centered at $t=t_0$ with certain dominant frequency $f_0$ and amplitude $A$, given by the following expression
	\begin{align}
		f(t) = A \exp \left(-\omega_0^2\frac{(t-t_0)^2}{2}\right)   \exp \left(\imag n \omega_0 t\right), \quad \text{with} \quad n\in\mathbb{R}^+ \text{and} \quad  \omega_0 = 2 \pi f_0 .
		\label{eq.complex_source_time_function_Gaussian}
	\end{align}
\end{enumerate}

The difference between a Gaussian (eq. \eqref{eq.source_time_function_Gaussian}) and a modulated Gaussian (eq. \eqref{eq.complex_source_time_function_Gaussian}) is that, in this study, the word \textit{modulated} means: a pulse that oscillates in time. The Gaussian envelope (eq. \eqref{eq.source_time_function_Gaussian}) determines the time duration of the wave packet and the oscillatory term (complex exponential in (eq. \eqref{eq.complex_source_time_function_Gaussian}) represents the central frequency of oscillation, dictating how fast the Gaussian pulse oscillates in time.

\begin{figure}
	\begin{center}
		\includegraphics[width=1\textwidth]{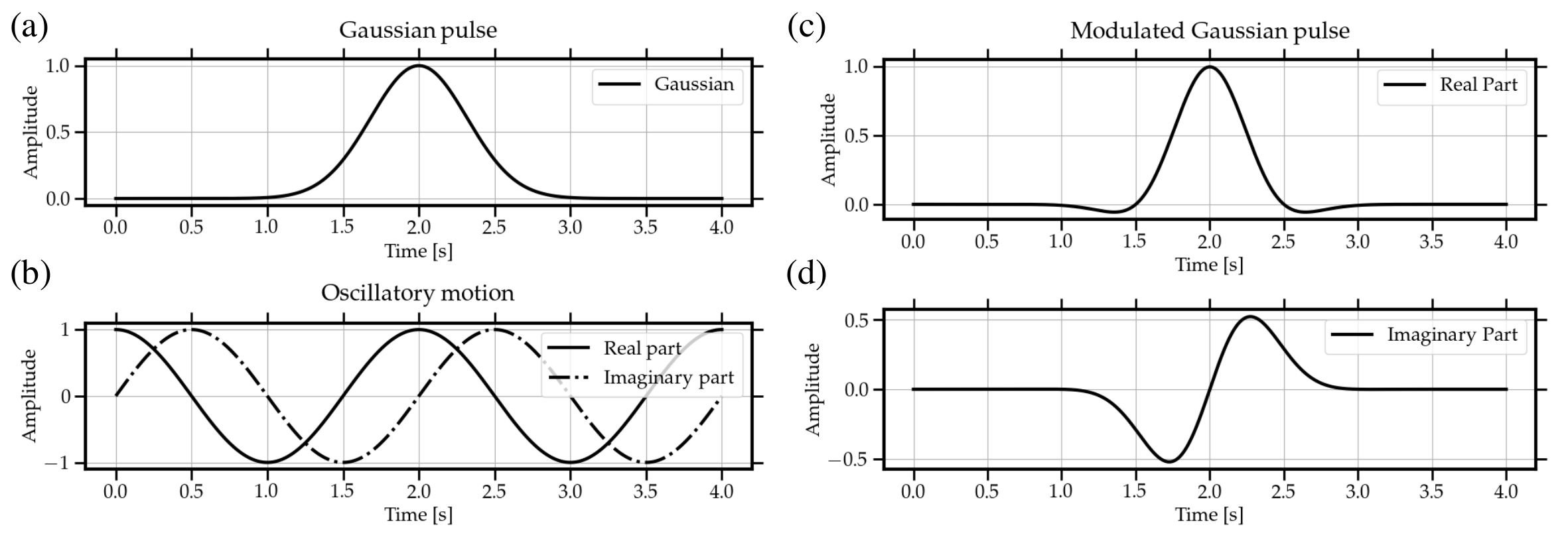}
		\caption{(a) Gaussian pulse (eq. \eqref{eq.source_time_function_Gaussian}) centered at $t_0=2$s and with a dominant period of 2s ($f_0=0.5$Hz). (b) The real and imaginary parts of the oscillatory term of the modulated Gaussian pulse (imaginary exponential in eq. \eqref{eq.complex_source_time_function_Gaussian}) with the same dominant frequency of $f_0=0.5$Hz. (c) Real and (d) Imaginary parts of the modulated Gaussian pulse (eq. \eqref{eq.complex_source_time_function_Gaussian}).}
		\label{Fig.Gaussian_pulses}
	\end{center}
\end{figure}

To develop some intuition, let us  we assume a Gaussian pulse centered at $t_0=2$s and with a dominant period of 2s ($f_0=0.5$Hz, see Fig. \ref{Fig.Gaussian_pulses}--a). The oscillatory term is also chosen to be with the same dominant frequency ($f_0=0.5$Hz, see Fig. \ref{Fig.Gaussian_pulses}--b). The resulting real part of the modulated Gaussian pulse does not resemble anymore a Gaussian like wavelet (see Fig. \ref{Fig.Gaussian_pulses}--c) and the imaginary part resembles the first-order derivative of a Gaussian (see Fig. \ref{Fig.Gaussian_pulses}--d). 

\subsection{Fr\'echet interaction density of the second-order wave equation}

We only analyze the Fr\'echet interaction density (eq. \eqref{eq.Frechet_interaction}) since the rest of the Fr\'echet kernels can be derived from this one, and have been previously analyzed in the literature for simple cases such as the second-order wave equation \citep[e.g.][]{fichtner2010book,menke2012geophysical}. For simplification purposes, we next assume that we have no data to compare, i.e., setting $\d=0$ in eq. \eqref{eq.1D_elastic_adjoint_wave_equation}. The assumed simulation parameters are given in Table \ref{tb.material_parameters}. 
\begin{table}
		\caption{FD simulation Parameters for the second-order wave equation}
\begin{center}
	{\renewcommand{\arraystretch}{1.5} \begin{tabular}{ | c | c | c |c | c | c | c | c |}
		\hline Velocity [m/s] & $\rho$ [kg/m$^3$] & $L_x$ [m]  & $x_s$ [m] & $x_r$ [m] & $f_0$ [Hz] & $n_x$ & $c \Delta t/\Delta x$\\ 
		\hline 1500 & 1000 & 4000 & 2666  & 1000  & 10 & 1000 & 1 \\ \hline    
	\end{tabular}}
	\label{tb.material_parameters}
\end{center}
\end{table}

\begin{figure}
	\begin{center}
		\includegraphics[width=1\textwidth]{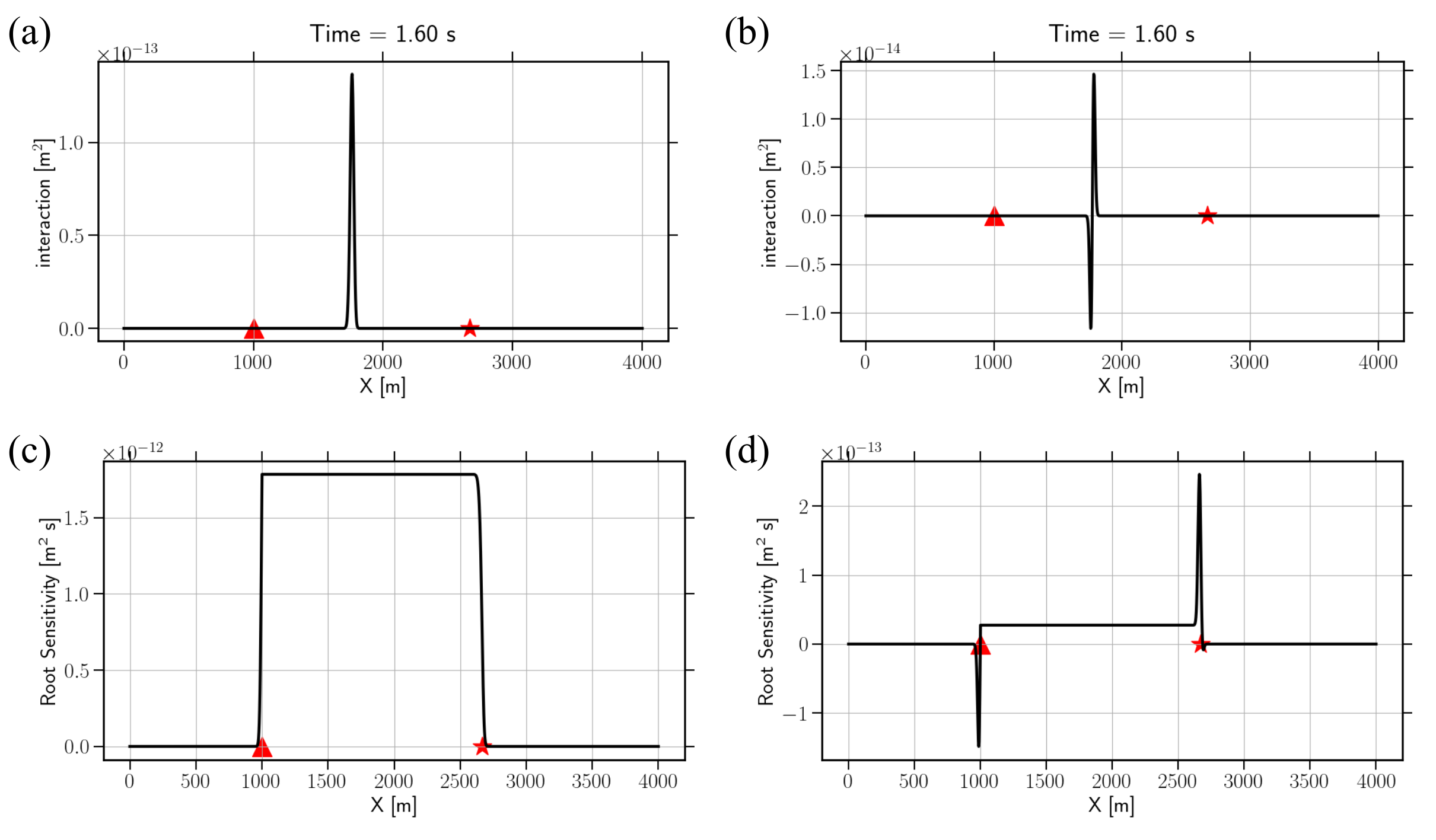}
		\caption{(a) and (b): Snapshots of the forward $u$ and adjoint $u^{\dagger}$ wavefield interactions $(u \,u^{\dagger})$ considering two different source time functions: (a) first-order and (b) second-order derivatives of a Gaussian (eq. \eqref{eq.source_time_function_Gaussian}). (c) and (d): Time integral of the Fr\'echet interaction densities corresponding to two different source time functions: (c) the first-order and (d) the second-order derivative of a Gaussian. Source and receiver locations are displayed by a star and a triangle respectively.}
		\label{Fig.Wave_Equation_Root_Kernels}
	\end{center}
\end{figure}

We first analyze the Fr\'echet interaction density (eq. \eqref{eq.Frechet_interaction}), considering that the source time functions are given by: (1) the first-order and (2) the second-order derivative of a Gaussian (eq. \eqref{eq.source_time_function_Gaussian}). Results, for the displayed time level, are shown in Fig.  \ref{Fig.Wave_Equation_Root_Kernels}--a, \ref{Fig.Wave_Equation_Root_Kernels}--b, where we can observe that the interaction $(u\, u^{\dagger})$ in both cases resembles the integral of the selected source time function. 

The corresponding time integrals of the Fr\'echet interaction densities are shown in Fig. \ref{Fig.Wave_Equation_Root_Kernels}--c, \ref{Fig.Wave_Equation_Root_Kernels}--d. We can observe that, as one can expect, the sensitivity between the source (star) and the receiver (triangle) is constant, since the medium is homogeneous. The only observable difference (apart from the amplitude) is the different sensitivity waveform located at the source and receiver locations. This is a well known feature when studying general Fr\'echet sensitivity kernels \citep[e.g.][]{fichtner2010book,menke2012geophysical} and it will always depend on the type of the source time function chosen for the forward and adjoint wavefields. 

The main observation of this section is that the time integral of the Fr\'echet interaction density between the source and receiver is always constant when we assume different source time functions for the forward $u$ and adjoint $u^{\dagger}$ wavefields, however, the interaction waveform $(u \, u^{\dagger})$ will vary for each different case.

\subsection{Fr\'echet interaction density of the Euler-Bernoulli beam equation}

Beams are 3D dimensional structural elements capable to tolerate perpendicular load primarily by resisting to axial bending \citep{watts2001isostasy}. In the Earth sciences, the Euler-Bernoulli beam equation is therefore used to model deformations of the Earth's lithosphere \citep{watts2015crustal,turcotte2002geodynamics,watts2001isostasy,burov2011rheology,watts2003lithospheric,chase1988flexural,steinberg2014flexural,jaeger2012elasticity} in response to forces such as the weight of mountains, the pull of mantle convection, or the bending due to oceanic or continental loading \citep{turcotte2002geodynamics}.

\begin{figure}
	\begin{center}
		\includegraphics[width=0.5\textwidth]{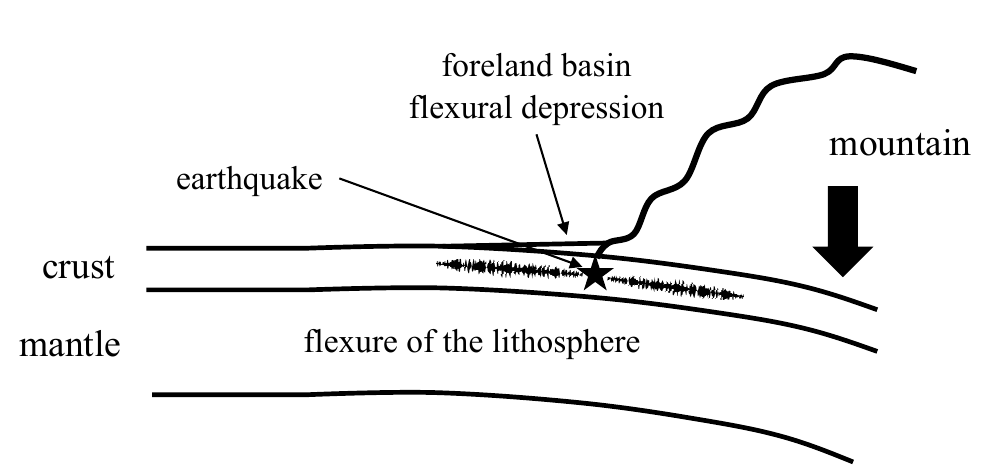}
		\caption{Two-dimensional sketch of lithospheric bending due to the weight of a mountain and transient deformation produced by an earthquake.}
		\label{Fig.Beam_Lithosphere}
	\end{center}
\end{figure}

We will next use the Euler-Bernoulli beam equation to model transient elastic deformation in the lithosphere due to an earthquake (see Fig. \ref{Fig.Beam_Lithosphere}). Material parameters are given in Table \ref{tb.Euler_Bernoulli_material_parameters}, were for the calculation of the moment of inertia $I$, we have considered the lithosphere as a rectangular cross-section. The source time function that generates elastic deformation is considered to be the real part of a modulated Gaussian pulse with a dominant frequency of 20 Hz (with $n=8$, see eq. \eqref{eq.complex_source_time_function_Gaussian}). 
\begin{table}
	\caption{FD simulation Parameters for the Euler-Bernoulli beam equation.}
	\begin{center}
		\resizebox{\textwidth}{!}{%
			{\renewcommand{\arraystretch}{1.5} 
				\begin{tabular}{ | c | c | c | c | c | c | c | c | c | c | c |}
					\hline
					$\rho$ [kg/m$^3$] & $E$ [GPa] & $A$ [m$^2$]  & $I$ [m$^4$] & $L_x$ [km] & width [km] & $x_s$ [m] & $x_r$ [m] & $f_0$ [Hz] & $n_x$ & $\sqrt{EI/\rho A}\Delta t/\Delta x^2$\\ 
					\hline 
					3000 & 70 & $10^{10}$ & $3\times 10^{18}$  & 300  & 100 & $2/3L_x$ & $1/4L_x$ & 20 & 1000 & 0.5 \\ 
					\hline    
			\end{tabular}}%
		\label{tb.Euler_Bernoulli_material_parameters}
		}
	\end{center}
\end{table}

\begin{figure}
	\begin{center}
		\includegraphics[width=1\textwidth]{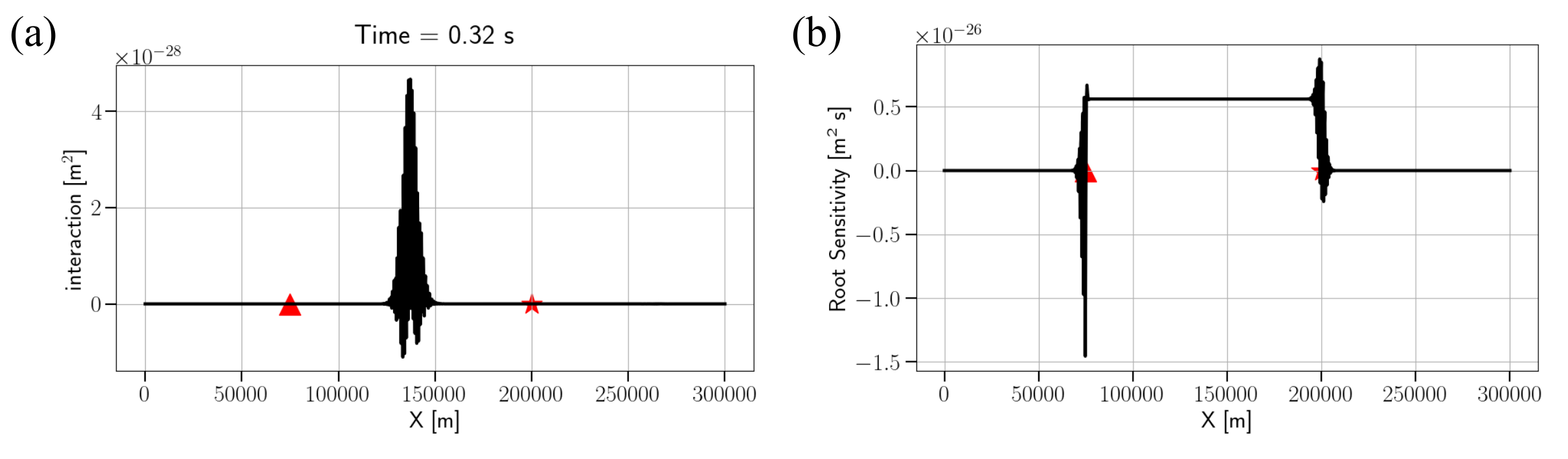}
		\caption{Euler-Bernoulli beam equation: (a) Snapshot of the forward $u$ and adjoint $u^{\dagger}$ wavefield interaction $(u \, u^{\dagger})$ at the specified time. (b) Corresponding time integral of the interaction density. Source and receiver locations are displayed by a star and a triangle respectively.}
		\label{Fig.Euler_Bornoulli_Root_Kernels}
	\end{center}
\end{figure}

Results are shown in Fig. \ref{Fig.Euler_Bornoulli_Root_Kernels}, where we can observe that the interaction between the forward $u$ and adjoint $u^{\dagger}$ wavefields $(u \, u^{\dagger})$ do not resemble anymore a clear wavelet (Fig. \ref{Fig.Euler_Bornoulli_Root_Kernels}--a), like in the case of the second-order wave equation (see Fig. \ref{Fig.Wave_Equation_Root_Kernels}). On the contrary, the Fr\'echet interaction density $u \, u^{\dagger}$ shows a high level of dispersion, which may not be a surprising result, since the Euler-Bernoulli beam equation is a dispersive PDE, meaning that different frequencies travel at different speeds (the phase velocity depends on frequency).

On the contrary, the time integral of the Fr\'echet interaction density shown in Fig. \ref{Fig.Euler_Bornoulli_Root_Kernels}–b exhibits a constant sensitivity between the source and the receiver, which aligns with the Fr\'echet sensitivity observed for the second-order wave equation (see Fig. \ref{Fig.Wave_Equation_Root_Kernels}). This, once again, is physically sound since we are considering a homogeneous medium.

\subsection{Fr\'echet interaction density of the complex transport equation}

As previously explained, the complex transport equation defined by eq. \eqref{eq.1D_imaginary_wave_equation} simply describes the propagation of the real and imaginary parts of the initial condition of motion by following the conventional second-order wave equation for each part (eq. \eqref{eq.1D_wave_equation}). This simply allows us to consider a complex valued initial condition of motion which, as we will next see, has interesting properties for the adjoint problem.

We consider that the source of motion is given by a modulated Gaussian pulse with a dominant frequency of 10 Hz ($n=8$, see eq. \eqref{eq.complex_source_time_function_Gaussian}). The assumed simulation parameters are those given in Table \ref{tb.material_parameters}. To compute the Fr\'echet interaction density between the forward $\u$ and adjoint $\u^{\dagger}$ wavefields $(\u^{\dagger} \, \u$), we explore the possibility of two cases for the selection of the adjoint source: (1) the adjoint source is given by a Dirac delta distribution located at the arrival time the forward wave and (2) the complex conjugate $\bar{\u}$ of the recorded wavefield. 

\begin{figure}
	\begin{center}
		\includegraphics[width=1\textwidth]{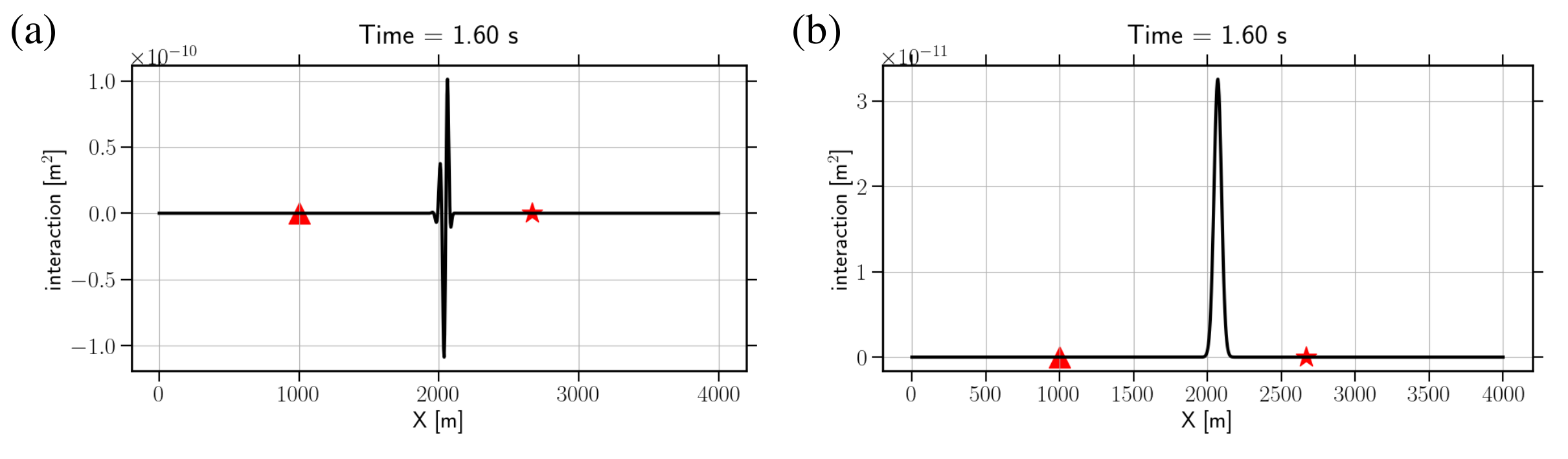}
		\caption{Snapshots of the forward $\u$ and adjoint $\u^{\dagger}$ wavefield Fr\'echet interaction density $(\u \, \u^{\dagger})$ considering two different source time functions for the adjoint wavefield: (a) a Dirac delta pulse at the arrival time of the forward wavefield at the receiver location and (b) the complex conjugate $\bar{\u}$ of the forward wavefield. Source and receiver locations are displayed by a star and a triangle respectively.}
		\label{Fig.Particles_ComplexWave_Root_Kernels}
	\end{center}
\end{figure}

Figure \ref{Fig.Particles_ComplexWave_Root_Kernels} shows the obtained results. For the first case (Figure \ref{Fig.Particles_ComplexWave_Root_Kernels}--a) we can observe that the interaction resembles simply a modified wavelet of the initial Gaussian condition. On the contrary, the second case when the adjoint source is equal to the complex conjugate of the forward displacement $(\u^{\dagger}=\bar{\u})$ the Fr\'echet interaction density $(\u^{\dagger}=\bar{\u})$ resembles a Gaussian waveform as one can expect.

\begin{figure}
	\begin{center}
		\includegraphics[width=1\textwidth]{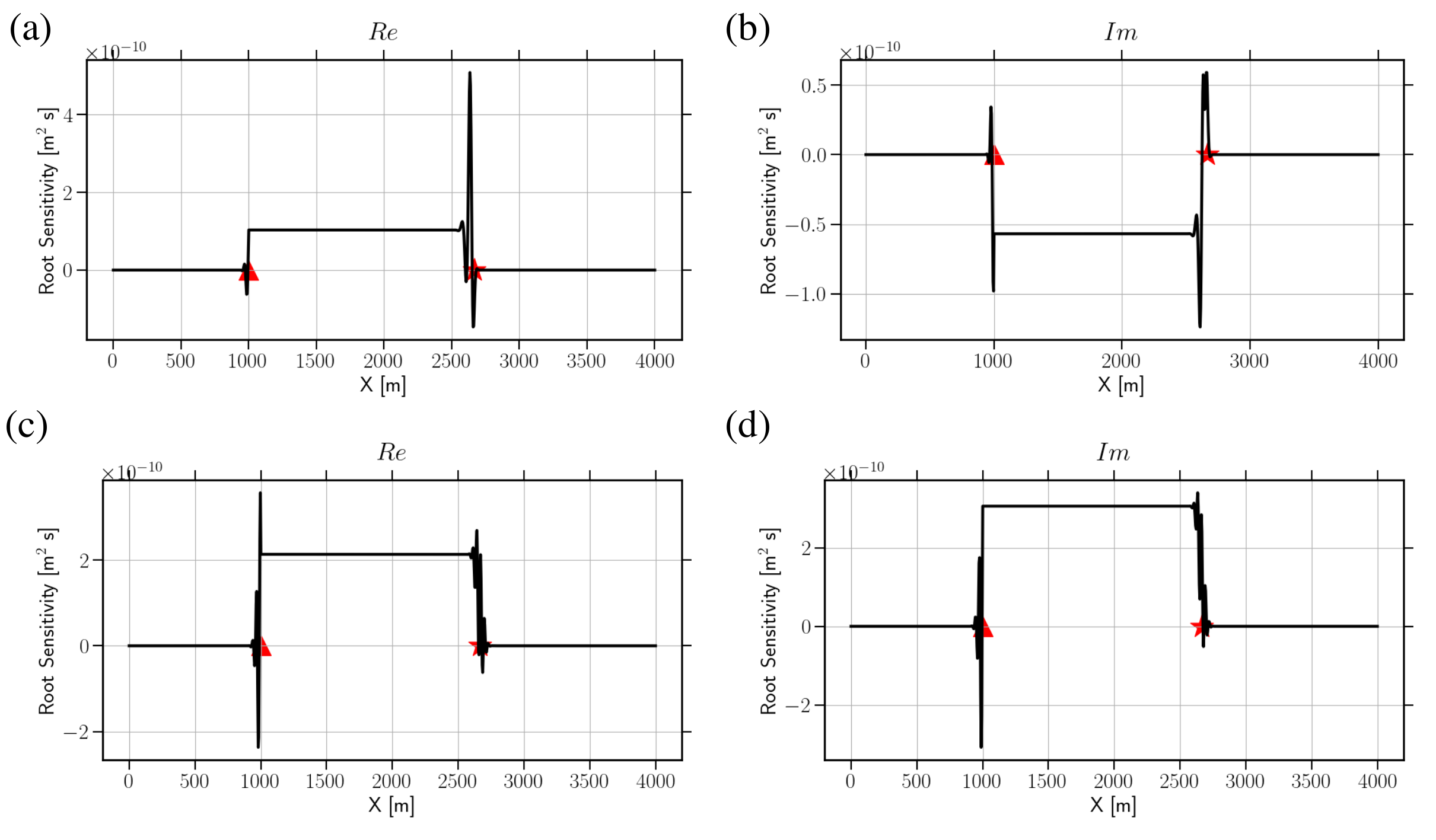}
		\caption{Time integral of the Fr\'echet interaction density considering two different source time functions for the adjoint wavefield of the complex-transport wave equation: (a)--(b) a Dirac delta pulse at the arrival time of the forward wavefield at the receiver location and (c)--(d) the complex conjugate $\bar{\u}$ of the forward wavefield. Source and receiver locations are displayed by a star and a triangle respectively.}
		\label{Fig.ComplexWave_Root_Kernels}
	\end{center}
\end{figure}

The corresponding time integrals of the Fr\'echet interaction densities are shown in Fig. \ref{Fig.ComplexWave_Root_Kernels}. We can observe that, as expected (because we are considering constant parameters), the sensitivity for the real and imaginary parts in all cases is simply a constant between the source and the receiver locations. The only differences, besides amplitudes, are the waveforms observed at the source and receiver locations.

\subsection{Fr\'echet interaction density of the (zero potential) Schr\"odinger equation}

The previous examples gave us the necessary ingredients to understand the case for the Schr\"odinger equation with zero potential as we will next see. First note that the decoupled Schr\"odinger equation has the same mathematical structure as the Euler-Bernoulli Beam equation (eq. \eqref{eq.Euler_Bernoulli_Beam}), thus real and imaginary parts of the Fr\'echet interaction densities can independently be computed by separating the initial condition of motion in its real and imaginary parts. 

For the simplification of simulation parameters, we choose atomic units [a.u.] \citep{mcquarrie1997physical}, i.e., $\hbar = m = T= 1$ with respect to the parameters given in Table \ref{tb.Schroedinger_material_parameters}. We consider that the source of motion is given by a modulated Gaussian pulse with a dominant atomic frequency of $1\times10^4$  [a.u. 1/s] (with $n=8$, see eq. \eqref{eq.complex_source_time_function_Gaussian}). 
\begin{table}
	\caption{FD simulation Parameters for the zero potential Schr\"odinger equation.}
	\begin{center}
		\resizebox{\textwidth}{!}{%
			{\renewcommand{\arraystretch}{1.5} 
				\begin{tabular}{ | c | c | c | c | c | c | c | c | c | c | c | c |}
					\hline
					$m$ [kg] & $\hbar$ [m$^2$ kg / s] & $L_x$ [m] &  $x_s$ [m] & $x_r$ [m] & $t$ [s] & $f_0$ [a.u. 1/s] & $n_x$ & $\hbar/2m\Delta t/\Delta x$\\ 
					\hline 
					$9.1\times 10^{-31}$ & $6.6\times 10^{-34}$  & $5.29\times 10^{-11}$ & $2/3L_x$ & $1/4L_x$ & $2.4\times10^{-17}$ [s] & $1\times10^{4}$ & 1000 & 0.5 \\ 
					\hline    
			\end{tabular}}%
			\label{tb.Schroedinger_material_parameters}
		}
	\end{center}
\end{table}

\begin{figure}
	\begin{center}
		\includegraphics[width=1\textwidth]{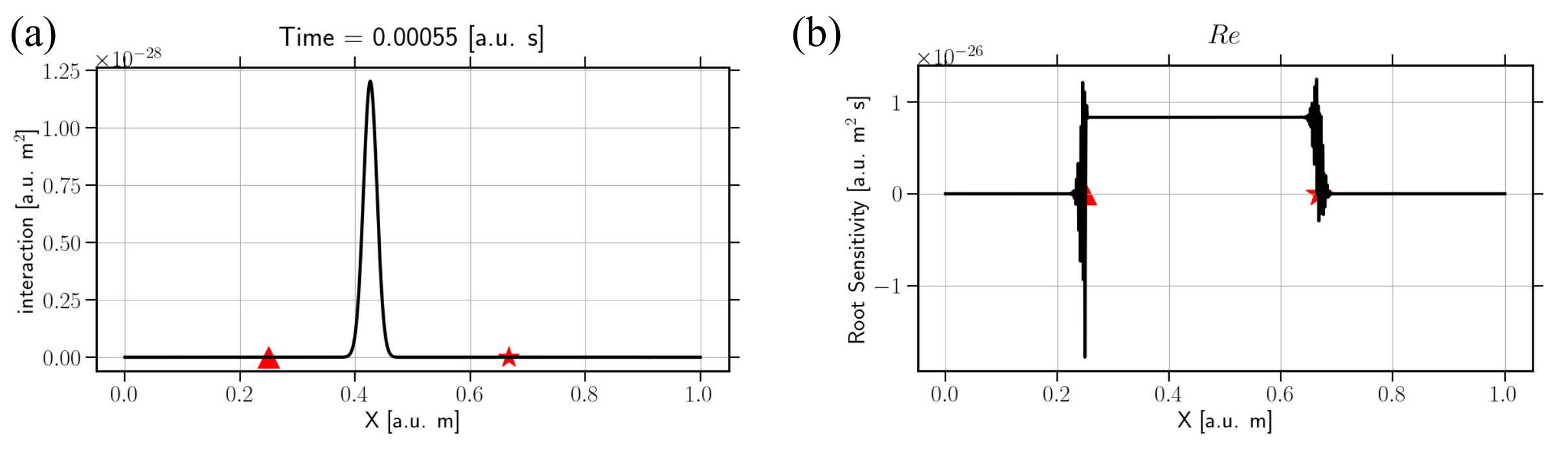}
		\caption{(a) Snapshot of the forward $\psi$ and adjoint $\psi^{\dagger}$ wavefunction interactions $(\psi\psi^{\dagger})$ considering that the source time functions for the adjoint wavefunction is the complex conjugate $\bar{\psi}$ of the recorded wavefunction. (b) Corresponding time integral of the Fr\'echet interaction density. Source and receiver locations are displayed by a star and a triangle respectively.}
		\label{Fig.Schroedinger_Root_Kernels}
	\end{center}
\end{figure}

For the selection of the adjoint source we simply choose the case when it is equal to the complex conjugate of the recorded forward wavefield $(\u^{\dagger}=\bar{\u})$. Results are presented in Fig. \ref{Fig.Schroedinger_Root_Kernels}, where we can observe that the Fr\'echet interaction density $(\u^{\dagger} \, \bar{\u}$ resembles a Gaussian as one can expect (Fig. \ref{Fig.Schroedinger_Root_Kernels}--a) and the time integral of the Fr\'echet interaction density is purely real with constant sensitivity between the source and receiver.

\section{Discussion}

\subsection{The Fr\'echet sensitivity kernels}

Fr\'echet sensitivity kernels quantify how parameter changes (of the chosen PDE) affect the corresponding outputs. Despite different Fr\'echet sensitivity kernels can be derived depending on the PDE, all include interactions between the a modified version of the forward and adjoint wavefields. The definition of the adjoint wavefield depends on the misfit functional that we choose to evaluate. For the time-dependent PDE studied here, the adjoint wavefield is a backward in time propagated wavefield, that follows the same physical rules of the forward wavefield (same PDE). After the selection of the misfit functional and calculation of the adjoint wavefield, we are able to have access to the different Fr\'echet sensitivity kernels.

\subsection{The Fr\'echet interaction densities}

We have defined the Fr\'echet interaction density as the simplest interaction between forward and adjoint wavefields. Based on this definition, we have analyzed the interaction density obtained for four different PDEs: the first group of two PDEs are real valued and the second group complex valued. 

For the group of real valued PDEs, we have considered: (i) second-order wave and (ii) the Euler-Bernoulli beam equations. For the complex-valued PDEs we have considered: (i) the complex-transport and (ii) the Schr\"odinger (with zero potential) equations. When we decompose both complex valued PDEs into real and imaginary parts, both simply correspond to the mentioned real valued PDEs. Thus, understanding Fr\'echet interaction densities for the real valued scenarios gives us the fundamentals to understand the complex valued scenarios. 

As one can expect, in all cases the Fr\'echet interaction density time integral is constant when constant material parameters are chosen. Interestingly, in all cases, depending of the initial condition of motion chosen, different sensitivity (in waveform and amplitude) is evidenced at the source and receiver locations.

In order to understand this effect at the source and receiver locations, we analyzed the Fr\'echet interaction density by separate for each considered PDE. We found that the interaction density travels like a wave (propagated by the considered PDE) with a waveform depending on the chosen initial condition.

While for real-valued PDEs, this may not seem a striking result, for complex-valued PDEs it is. This is because we are free to choose the source of motion that generates the adjoint wavefield. Thus for the particular selection that the source of motion of the adjoint wavefield is the complex conjugate of the recorded forward wavefield we recover always a Gaussian like shape for the Fr\'echet interaction density. This corresponds to the well known Born rule of quantum mechanics, recovered as a particular Fr\'echet interaction density.                 

\subsection{The Born rule as a Fr\'echet interaction density}

The Born rule is a fundamental principle in quantum mechanics that describes a mathematical relationship between the wavefunction and measurement outcomes. It is essential for understanding what we observe using the Schr\"odinger equation \citep[e.g.][]{griffiths2018introduction,shankar2012principles,zee2010quantum,sakurai2020modern}. The Born rule states that the probability \( P(\x ) \) of measuring a particle at position \( \x \) is given by the modulus of the wavefunction as follows
\begin{align}
	P(\x,t) = |\WF(\x,t))|^2 =  \WF(\x,t)\,\overline{\WF}(\x,t) , 
	\label{eq.Wave_Function_Collapse}
\end{align}
where $\overline{\WF}$ is the complex conjugate of the wavefunction $(\WF)$. Note that the wave function is a complex valued vector, but the square modulus, \( |\WF(\x)|^2 \), gives a real, non-negative probability density for finding the particle at a specific location. The Born rule provides the link between before and after measurements in quantum mechanics. The collapse of the wavefunction is, however, still not yet fully understood.

Although the Born rule has never been derived/related from Fr\'echet sensitivity kernels, both concepts can be viewed as addressing a form of sensitivity: the Born rule characterizes the sensitivity of measurement outcomes, while Fr\'echet sensitivity kernels quantify how a functional responds to perturbations in the parameters on which it depends. Both therefore reflect a common abstract idea—how small changes in a state influence the observable behavior of a system—even though they arise from distinct mathematical and physical frameworks.

By applying the adjoint method to the Schr\"odinger equation and computing the associated Fr\'echet sensitivity kernel and interaction density, we find that the resulting sensitivity of a measurement (i.e., the probability density) to perturbations in the wavefunction takes exactly the same form as the Born rule. Specifically, the Fr\'echet interaction density for the Schr\"odinger equation describes how infinitesimal changes in the wavefunction influence the probability distribution, and for position measurements this sensitivity coincides with the squared modulus of the wavefunction.

\section{Conclusions}

We extended the adjoint method to complex-valued PDEs and applied it to four representative equations. We introduced the Fr\'echet interaction density, defined as the instantaneous overlap between the forward and adjoint wavefields, from which Fr\'echet kernels can be constructed.

We demonstrated how this framework clarifies the structure of Fr\'echet sensitivity kernels for both real- and complex-valued PDEs. In all considered cases with constant material parameters, the time integral of the Fr\'echet interaction density is constant between the source of motion and the location of measurement. However, the Fr\'echet interaction density waveform propagates as a wave governed by the underlying PDE, with a waveform determined by the chosen initial condition of motion. While this behavior is expected for real-valued PDEs, its implications for complex-valued PDEs are particularly interesting: for the Schr\"odinger and complex-transport equations, choosing the adjoint source as the complex conjugate of the forward wavefield leads to a Gaussian-shaped instantaneous sensitivity.

We further established a direct connection between the Born rule of quantum mechanics and the Fr\'echet interaction density of the Schr\"odinger equation with zero potential.
 Our analysis suggests that the Born rule can be viewed, within this framework, as a particular positive-definite sensitivity density, that it is naturally related to the sensitivity of quantum systems to perturbations of the wavefunction. Under this interpretation, a measurement corresponds to the response of the system to small changes in the model parameters, and the resulting sensitivity interaction reproduces the probability density given by the Born rule.

This connection between Fr\'echet interaction densities and the Born rule may provide a different foundation for a more unified understanding of quantum measurements, linking concepts from quantum theory with adjoint-based sensitivity analysis commonly used in optimization and inverse problems.

\section{Acknowledgments}

Numerical computations were performed on the S-CAPAD/DANTE platform, IPGP, France.

\section*{Conflict of interest}

The authors declare that they have no conflict of interest.

\section*{Funding}

C.N. acknowledges financial support from the Institut de Physique du Globe de Paris (IPGP). No additional external funding was received for this work.

\section*{Data availability}

The numerical simulations presented in this work were generated using finite-difference codes. The data and scripts supporting the findings of this study are available from the corresponding author upon reasonable request.

\footnotesize
\bibliographystyle{apalike}
\bibliography{Biblio}

@article{tonti1973variational,
  title={On the variational formulation for linear initial value problems},
  author={Tonti, Enzo},
  journal={Annali di Matematica Pura ed Applicata},
  volume={95},
  number={1},
  pages={331--359},
  year={1973},
  publisher={Springer}
}

@book{Morse1953,
	title={{Methods of Theoretical Physics}},
	author={Philip Morse and Herman Feshbach},
	year={1953},
	volume={1},
	publisher={McGraw Hill}
}

@book{igel2017computational,
	title={Computational Seismology: A Practical Introduction},
	author={Igel, Heiner},
	year={2017},
	publisher={Oxford University Press}
}

@book{moczo2014finite,
	title={The finite-difference modelling of earthquake motions: Waves and ruptures},
	author={Moczo, Peter and Kristek, Jozef and G{\'a}lis, Martin},
	year={2014},
	publisher={Cambridge University Press}
}

@book{fichtner2010book,
	title={Full seismic waveform modelling and inversion},
	author={Fichtner, Andreas},
	year={2010},
	publisher={Springer}
}

@book{griffel2002applied,
	title={Applied Functional Analysis},
	author={Griffel, David},
	year={1981},
	publisher={Wiley}
}

@book{boyd2004convex,
  title={Convex Optimization},
  author={Boyd, Stephen and Boyd, Stephen P and Vandenberghe, Lieven},
  year={2004},
  publisher={Cambridge University Press}
}

@article{frechet1912notion,
	title={Sur la notion de diff{\'e}rentielle totale},
	author={Fr{\'e}chet, Maurice},
	journal={Nouvelles Annales de Math{\'e}matiques},
	volume={12},
	pages={385--403},
	year={1912}
}

@article{frechet1925notion,
	title={La notion de diff{\'e}rentielle dans l'analyse g{\'e}n{\'e}rale},
	author={Fr{\'e}chet, Maurice},
	journal={Annales Scientifiques de l'{\'E}cole Normale Sup{\'e}rieure},
	volume={42},
	pages={293--323},
	year={1925}
}

@article{frechet1911notion,
	title={Sur la notion de diff{\'e}rentielle},
	author={Fr{\'e}chet, Maurice},
	journal={Comptes Rendus de l'Acad\'mie Sciences Paris},
	volume={152},
	pages={845--847},
	year={1911}
}

@book{zeidler1984nonlinear,
	title={{Nonlinear functional analysis and its applications III: variational methods and optimization}},
	author={Zeidler, Eberhard},
	year={1984},
	publisher={Springer}
}

@book{zeidler1990nonlinear,
	title={{Nonlinear functional analysis and its applications II: linear monotone operators}},
	author={Zeidler, Eberhard},
	year={1990},
	publisher={Springer}
}

@book{menke2012geophysical,
	title={{Geophysical Data Analysis: Discrete Inverse Theory}},
	author={Menke, William},
	year={2012},
	publisher={Academic Press}
}

@article{abreu2024understanding,
	title={Understanding the Adjoint Method in Seismology: Theory and Implementation in the Time Domain},
	author={Abreu, Rafael},
	journal={Surveys in Geophysics},
	volume={45},
	number={5},
	pages={1363--1434},
	year={2024},
	publisher={Springer}
}

@book{arfken2011mathematical,
	title={Mathematical methods for physicists: a comprehensive guide},
	author={Arfken, George B and Weber, Hans J and Harris, Frank E},
	year={2011},
	Edition={Seventh},
	publisher={Elsevier}
}

@book{farlow1993partial,
	title={Partial differential equations for scientists and engineers},
	author={Farlow, Stanley J},
	year={1993},
	publisher={Dover}
}

@article{hanson1981sufficiency,
  title={{On sufficiency of the Kuhn-Tucker conditions}},
  author={Hanson, Morgan A},
  journal={J. Math. Anal. Appl},
  volume={80},
  number={2},
  pages={545--550},
  year={1981}
}

@article{hanson1999invexity,
	title={{Invexity and the Kuhn--Tucker theorem}},
	author={Hanson, Morgan A},
	journal={Journal of mathematical analysis and applications},
	volume={236},
	number={2},
	pages={594--604},
	year={1999},
	publisher={Elsevier}
}

@book{griffiths2018introduction,
	title={Introduction to quantum mechanics},
	author={Griffiths, David J and Schroeter, Darrell F},
	year={2018},
	publisher={Cambridge University Press}
}

@book{shankar2012principles,
	title={Principles of quantum mechanics},
	author={Shankar, Ramamurti},
	year={2012},
	publisher={Springer}
}

@book{zee2010quantum,
	title={Quantum field theory in a nutshell},
	author={Zee, Anthony},
	volume={7},
	year={2010},
	publisher={Princeton University Press}
}

@book{sakurai2020modern,
	title={Modern quantum mechanics},
	author={Sakurai, Jun John and Napolitano, Jim},
	year={2020},
	publisher={Cambridge University Press}
}

@book{karnopp2012system,
	title={System dynamics: modeling, simulation, and control of mechatronic systems},
	author={Karnopp, Dean C and Margolis, Donald L and Rosenberg, Ronald C},
	year={2012},
	publisher={John Wiley \& Sons}
}

@Book{            Timoshenko_1951aa,
	author        = {Timoshenko, S. and Goodier, J.N.},
	booktitle     = {Theory of Elasticity},
	edition       = {2},
	langid        = {english},
	publisher     = {McGraw-Hill Book Company},
	title         = {Theory of Elasticity},
	year          = {1951}
}

@book{watts2001isostasy,
	title={{Isostasy and Flexure of the Lithosphere}},
	author={Watts, Anthony Brian},
	year={2001},
	publisher={Cambridge University Press}
}

@incollection{watts2015crustal,
	year={2015},
	booktitle={Treatise on Geophysics - Volume 6},
	pages = {337–348},
	editor={Gerard Schubert},
	title={Crustal and lithosphere dynamics: an introduction and overview},
	author={Watts, A},
	publisher={Elsevier}
}

@book{turcotte2002geodynamics,
	title={Geodynamics},
	author={Turcotte, Donald L and Schubert, Gerald},
	year={2002},
	publisher={Cambridge University Press}
}

@article{burov2011rheology,
	title={Rheology and strength of the lithosphere},
	author={Burov, Evgene B},
	journal={Marine and Petroleum Geology},
	volume={28},
	number={8},
	pages={1402--1443},
	year={2011},
	publisher={Elsevier}
}

@article{watts2003lithospheric,
	title={Lithospheric strength and its relationship to the elastic and seismogenic layer thickness},
	author={Watts, AB and Burov, EB},
	journal={Earth and Planetary Science Letters},
	volume={213},
	number={1-2},
	pages={113--131},
	year={2003},
	publisher={Elsevier}
}

@article{chase1988flexural,
	title={{Flexural isostasy and uplift of the Sierra Nevada of California}},
	author={Chase, Clement G and Wallace, Terry C},
	journal={Journal of Geophysical Research: Solid Earth},
	volume={93},
	number={B4},
	pages={2795--2802},
	year={1988},
	publisher={Wiley Online Library}
}

@article{steinberg2014flexural,
	title={{Flexural response of a continental margin to sedimentary loading and lithospheric rupturing: The mountain ridge between the Levant basin and the Dead Sea transform}},
	author={Steinberg, J and Gvirtzman, Z and Garfunkel, Z},
	journal={Tectonics},
	volume={33},
	number={2},
	pages={166--186},
	year={2014},
	publisher={Wiley Online Library}
}

@book{jaeger2012elasticity,
	title={Elasticity, Fracture and Flow: with Engineering and Geological Applications},
	author={Jaeger, John Conrad},
	year={2012},
	publisher={Springer Science \& Business Media}
}

@book{mcquarrie1997physical,
	title={Physical chemistry: a molecular approach},
	author={McQuarrie, Donald Allan and Simon, John Douglas},
	volume={1},
	year={1997},
	publisher={University science books Sausalito, CA}
}

\end{document}